\documentclass[
	reprint,
	superscriptaddress,
	showkeys,
	aps,
	pra,
	longbibliography,
	nofootinbib
]{revtex4-2}

\usepackage[utf8]{inputenc}

\usepackage{xcolor}
\definecolor{blau}{cmyk}{1.0,0.2,0.0,0.4}
\definecolor{rot}{cmyk}{0.04,1.0,0.8,0.07}
\definecolor{purple}{cmyk}{0.08,1.0,0.3,0.36}

\usepackage[
	colorlinks,
	pdfpagelabels,
	breaklinks,
	pdfstartview=FitH,
	bookmarksopen=true,
	bookmarksnumbered=true,
	bookmarksopenlevel=2,
	plainpages=false,
	hypertexnames=false,
	citecolor=rot,
	linkcolor=blau,
	urlcolor=purple,
	pdftitle={Lattice MC meets lattice FRG},
	pdfauthor={Niklas Zorbach, Jan Philipp Klinger, Owe Philipsen, Jens Braun},
	pdfkeywords={non-perturbative methods, lattice simulations, FRG, symmetry breaking/restoration, Z(2) scalar field theory}
]{hyperref}

\usepackage{bm,dsfont}
\usepackage{tensor}
\usepackage{amsmath,amssymb,amsthm}
\usepackage{physics}
\usepackage{upgreek}
\usepackage[bb=boondox]{mathalfa}
\usepackage{mathtools}

\usepackage{bookmark}
\usepackage{graphicx}
\usepackage[caption=false]{subfig}
\usepackage{dcolumn}
\usepackage{comment}

\usepackage[utf8]{inputenc}
\usepackage{pgfplots}
\DeclareUnicodeCharacter{2212}{−}
\usepgfplotslibrary{groupplots,dateplot}
\usetikzlibrary{patterns,shapes.arrows}
\pgfplotsset{compat=newest}

\usepackage{tikz}

\usepackage{orcidlink}

\usepackage[acronym]{glossaries-extra}
\glssetcategoryattribute{acronym}{nohyperfirst}{true}
\setabbreviationstyle[acronym]{long-short}

\newacronym{FRG}{fRG}{functional Renormalization Group}
\newacronym{SSB}{SSB}{spontaneous symmetry breaking}
\newacronym{RG}{RG}{Renormalization Group}
\newacronym{UV}{UV}{ultraviolet}
\newacronym{IR}{IR}{infrared}
\newacronym{LPA}{LPA}{local potential approximation}
\newacronym{KT}{KT}{Kurganov-Tadmor}
\newacronym{LSD}{LSD}{lattice site decoupling}
\newacronym{MC}{MC}{Monte Carlo}
\newacronym{HMC}{HMC}{Hybrid Monte Carlo} 
\newacronym{QCD}{QCD}{Quantum Chromodynamics}

\allowdisplaybreaks

\newcommand\numberthis{\addtocounter{equation}{1}\tag{\theequation}}

\def\d{\,\mathrm{d}}

\usepackage[capitalise,sort,compress]{cleveref}
\Crefname{section}{Sec.}{Sections}
\Crefname{table}{Tab.}{Tables}

\begin{document}

\title{Lattice Monte Carlo meets lattice functional Renormalization Group:\\ A quantitative comparison}

\author{Niklas Zorbach~\orcidlink{0000-0002-8434-5641}}
\email{niklas.zorbach@tu-darmstadt.de}
\affiliation{
	Technische Universität Darmstadt, Department of Physics, Institut für Kernphysik, Theoriezentrum,\\
	Schlossgartenstra{\ss}e 2, D-64289 Darmstadt, Germany
}

\author{Jan Philipp Klinger~\orcidlink{0000-0002-8626-6322}}
\email{klinger@itp.uni-frankfurt.de}
\affiliation{Goethe-University Frankfurt am Main, Institute for Theoretical Physics,
	Max-von-Laue-Str. 1, 60438 Frankfurt am Main, Germany
}

\author{Owe Philipsen~\orcidlink{0000-0003-2046-7292}}
\email{philipsen@itp.uni-frankfurt.de}
\affiliation{Goethe-University Frankfurt am Main, Institute for Theoretical Physics,
	Max-von-Laue-Str. 1, 60438 Frankfurt am Main, Germany
}

\author{Jens Braun~\orcidlink{0000-0003-4655-9072}}
\email{jens.braun@tu-darmstadt.de}
\affiliation{
	Technische Universität Darmstadt, Department of Physics, Institut für Kernphysik, Theoriezentrum,\\
	Schlossgartenstra{\ss}e 2, D-64289 Darmstadt, Germany
}

\affiliation{Helmholtz Research Academy Hesse for FAIR, Campus Darmstadt, D-64289 Darmstadt, Germany}

\affiliation{ExtreMe Matter Institute EMMI, GSI, Planckstra{\ss}e 1, D-64291 Darmstadt, Germany}

\begin{abstract}
	Lattice Monte Carlo (MC) simulations and the functional Renormalization Group (RG) are powerful approaches that allow for quantitative studies of non-perturbative phenomena such as bound-state formation, spontaneous symmetry breaking and phase transitions. 
	While results from both methods have recently shown remarkable agreement for many observables, e.g., in Quantum Chromodynamics, an analysis of deviations in certain quantities turns out to be challenging. 
	This is because calculations with the two methods are based on different approximations, regularizations and scale fixing procedures. 
	In the present work, we present a framework for a more direct comparison by formulating the functional RG approach on a finite spacetime lattice. 
	This removes all ambiguities of regularization, finite size and scale fixing procedures in concrete studies. 
	By investigating the emergence of spontaneous symmetry breaking and  phase transitions in a $Z(2)$ scalar theory in $d=1,2,3$ spacetime dimensions, we demonstrate at the example of the local potential approximation how this framework can be used to evaluate and compare the systematic errors of both approaches.
\end{abstract}

\maketitle


%
\section{Introduction}\label{sec:introduction}
Phase transitions in strongly coupled systems are intensively studied in many areas of research,
and require non-perturbative methods to arrive at reliable theoretical predictions. 
One example is \gls{QCD}, whose phase structure is relevant for the early universe, heavy-ion collisions and neutron star physics.
In recent years, investigations 
at high temperature and low densities have been pushed to a new level with first-principles calculations, see, e.g., Refs.~\cite{HotQCD:2018pds,Bonati:2018nut,HotQCD:2019xnw,Bonati:2018fvg,Borsanyi:2020fev,wilson2,Kotov:2021rah,Kotov:2021hri,Cuteri_2021,Dini:2021hug,Cuteri:2022vwk,DAmbrosio:2022kig,Ding:2023oxy,Ding:2024sux,Klinger:2025xxb,Zhang:2025vns,Borsanyi:2025dyp} for recent lattice \gls{QCD} studies,  Refs.~\cite{Fu:2019hdw,Braun:2020ada,Gao:2020fbl,Gao:2021vsf,Gunkel:2021oya,Braun:2023qak,Fu:2023lcm,Bernhardt:2023hpr} for recent first-principles studies based on functional approaches, 
and Refs.~\cite{Fischer:2018sdj,Dupuis:2020fhh,Philipsen:2021qji,Aarts:2023vsf} for reviews. 
While both approaches are inherently non-perturbative, they have complementary systematic errors, strengths and weaknesses.
This motivates a systematic understanding of the former by detailed comparisons.

Although results from lattice \gls{QCD} calculations and first-principles functional studies show remarkable agreement for many observables, an analysis of the origin of differences remains difficult, due to the many possible sources, such as different implementations of the \gls{QCD} action and its symmetries, cutoff effects, finite-volume effects, and truncations. 
Moreover, different scale fixing procedures are often used in lattice \gls{MC} and functional studies which potentially results in a non-trivial matching procedure for the parameters of the theory under consideration. 
However, for a quantitative comparison and a rigorous understanding of the effect of approximations, it is necessary to eliminate any non-trivial matching procedure for parameters.

In this work we aim to overcome some of these problems by formulating the functional \gls{RG} approach on a finite spacetime lattice.
This allows to trivially relate the bare actions entering lattice \gls{MC} and \gls{FRG} studies, and in particular obviates a continuum limit
before a meaningful comparison.
As a first step, we restrict ourselves to a scalar theory without gauge degrees of freedom. This provides a useful framework for a quantitative analysis of the effect of a plethora of artefacts which are also present in \gls{QCD} studies, such as cutoff artefacts, finite-volume effects, and truncation artefacts. 
Artefacts associated with different fermion implementations in lattice simulations may in principle be analyzed within such a framework as well.
Moreover, the possibility of a clear comparison between the two methods is appealing as it may trigger a cross-fertilization with respect to improvements of both methods.

Based on earlier \gls{FRG} studies of quantum field theories in a finite spacetime volume~\cite{Braun:2004yk,Braun:2005gy,Braun:2009ruy,Fister:2015eca} and on a spacetime lattice~\cite{Dupuis:2008vr,Machado:2010wi,Rancon:2011aa,Krieg:2017hcm,Pawlowski:2017rhn}, we set up a framework for clean direct comparisons of lattice \gls{MC} and \gls{FRG} studies, which allows for a quantitative understanding of the effect of the approximations underlying these two methods. 
This is of particular relevance for \gls{QCD} applications but also beyond. 
For concreteness, we shall focus on a $Z(2)$ scalar field theory in $d=1,2,3$ spacetime dimensions in the present work since it is simulable with high
precision and allows for particularly clean comparisons of this kind. 
The consideration of different numbers of spacetime dimensions is interesting as it allows to directly test whether a non-trivial momentum dependence in correlation functions is indeed suppressed when the number of spacetime dimensions is increased. 
Because of dimensional reduction, the case of spin-type models in $d=3$ is of particular interest for \gls{QCD} at finite temperature. 
For example, $O(4)$-type models are expected to provide an effective description of the chiral \gls{QCD} phase transition at low densities and the restoration of the $Z(2)$ symmetry may play a prominent role close to a potentially existing critical endpoint in the \gls{QCD} phase diagram. 

This work is organized as follows: In Sec.~\ref{sec:model}, we introduce the concrete model for our numerical studies. 
The two methods considered in our present work, lattice \gls{MC} and lattice \gls{FRG}, are then discussed in Secs.~\ref{sec:MC-simulation-intro} and~\ref{sec:lfrg}, respectively. 
While we keep the general introduction of the lattice \gls{MC} approach brief, we provide a more detailed discussion of the lattice \gls{FRG} approach. 
In general, the latter provides us with a set of differential equations for correlation functions on a spacetime lattice. 
In addition to a discussion of regulator functions, general aspects of RG flows on a spacetime lattice, and the connection to the standard continuum \gls{FRG} approach, we discuss the truncation underlying our numerical calculations and show in which limits this truncation already provides us with exact results. 
To be specific, we consider the so-called \gls{LPA} in our numerical studies which corresponds to the leading-order approximation in terms of a derivative expansion of the effective action. 
Note that this truncation is the simplest truncation in the \gls{FRG} approach which already includes fluctuation effects. 
Our main results are presented in Sec.~\ref{sec:res}, where we also provide an intrinsic estimate of the uncertainties of \gls{LPA} by a direct computation of momentum-dependent corrections to the two-point function. In addition, we compare lattice \gls{MC} and \gls{FRG} results for the order parameter of our $Z(2)$ model and the susceptibility across lattices with different sizes. 
Our conclusions can be found in Sec.~\ref{sec:conclusion}.

\section{Model}
\label{sec:model}
We consider a single-component real scalar field $\phi$ on a $d$-dimensional isotropic lattice $\mathcal{V} = \{ \bm x = (x_1, \dots, x_d) \, | \, x_\mu = a n_\mu, \, n_\mu \in \{0, \dots, N_\mu-1\} \} \subset (a \mathbb{Z})^d $ with lattice spacing $a$ and periodic boundary conditions for $\phi$. The extent of the lattice is assumed to be the same in all directions, $N_\mu = N_\sigma$.
The partition function reads
\begin{align*}
	\numberthis\label{eq:partition-function}
	\mathcal{Z}[J] &= \int \mathcal{D} \phi \, e^{-S[\phi] + J \cdot \phi} \, ,
\end{align*}
where $J \cdot \phi = a^d \sum_{\bm x \in \mathcal{V}} J_{\bm x} \phi_{\bm x}$, $S[\phi] = S(\{\phi_{\bm x \in \mathcal{V}}\})$,\footnote{From here on, $A[\phi]$ is short for $A(\{\phi_{\bm x \in \mathcal{V}}\})$.} and the measure of the partition function is defined as
\begin{align*}
	\numberthis\label{eq:path-integral-measure}
	\int \mathcal{D} \phi = \prod_{\bm y \in \mathcal{V}} a^{d_\phi} \int^{\infty}_{-\infty} \d \phi_{\bm y} \,.
\end{align*}
Here, $d_\phi = (d-2)/2$ is the mass dimension of the field~$\phi$.
Note that, with this definition of the measure, the path integral is dimensionless. 

Furthermore, we work with a discretized bare action $S[\phi]$ of the following form
\begin{align*}
	\numberthis\label{eq:action-on-lattice}
	S[\phi] =  a^d \sum_{\bm x \in \mathcal{V}} \left[  \frac{1}{2}\sum^d_{\mu = 1}  \Delta_{\mu}^f \phi_{\bm x}\Delta_{\mu}^f\phi_{\bm x}  +    \mathcal{U}(\phi_{\bm x}) \right] \, ,
\end{align*}
where $\Delta_{\mu}^f \phi_{\bm x}= \left(\phi_{\bm x+\bm e_\mu}-\phi_{\bm x} \right)/a$ is the discretized forward derivative and $\mathcal{U}(\phi_{\bm x})$ denotes the bare potential of the form 
\begin{align*}
	\numberthis\label{eq:potential-on-lattice}
	\mathcal{U}(\phi_{\bm x}) = \frac{1}{2} m^2 \phi^2_{\bm x} + \frac{1}{4!} \lambda \phi^4_{\bm x} - c \phi_{\bm x}\,.
\end{align*}
Here, we introduced an external homogeneous field $c$ which couples linearly to the field $\phi$. 
For the quartic coupling we assume~$\lambda>0$.

By rewriting the kinetic part of the action~\eqref{eq:action-on-lattice} in momentum space, we find
\begin{align*}
	\numberthis\label{eq:kinetic-term-of-bare-action}
	S[\phi] = \frac{1}{V} \sum_{\bm q \in \tilde{\mathcal{V}}} \frac{\epsilon^2_{\bm q}}{2} \tilde \phi_{-\bm q} \tilde \phi_{\bm q} +    a^d \sum_{\bm x \in \mathcal{V}}\mathcal{U}(\phi_{\bm x}) \, ,
\end{align*}
where $V = (a N_{\sigma})^d$ denotes the volume of the system and $\tilde{\mathcal{V}} = \{ \bm q = (q_1, \dots, q_d) \, | \, q_\mu = \frac{2 \pi  n_\mu}{a N_\sigma} \, , \,n_\mu \in \{0, \dots N_\sigma-1\}\}$ the corresponding momentum space.
The kinetic energy~$\epsilon_{\bm q}$ is defined as
\begin{align*}
	\numberthis\label{eq:dispersion-relation}
	\epsilon^2_{\bm q} = \sum_{\mu = 1}^{d} \Bigg[\frac{2}{a} \sin(\frac{1}{2} a  q_\mu)\Bigg]^2 \, .
\end{align*}
This quantity determines the kinetic energy levels for a given lattice momentum~$\bm q$.
Note that the functional form of the kinetic energy reflects the periodic boundary conditions. 

\subsection{Spontaneous symmetry breaking}\label{sec:spontaneous-symmetry-breaking}
\Gls{SSB} of the global $Z(2)$ symmetry of our theory can be realized only in the thermodynamic limit, $N_\sigma\to \infty$ (for a fixed lattice spacing). 
In any finite volume quantum fluctuations inevitably restore the $Z(2)$ symmetry. 
From a mathematical standpoint, \gls{SSB} can be defined as a limiting process where an external $Z(2)$ symmetry breaking source (e.g., given in form of the parameter $c$ in our action above) is removed {\it after} the extrapolation to the infinite volume limit has been taken.  

An order parameter for spontaneous $Z(2)$ symmetry breaking is given by the ``magnetization", 
\begin{align*}
	\numberthis\label{eq:magnetization}
	\langle M \rangle := \lim_{c\to 0} \lim_{V\to \infty} \langle M \rangle_{V, c} \, ,
\end{align*}
where $M$ is the average field value
\begin{align*}
	\numberthis\label{eq:M}
	M = \frac{a^d}{V} \sum_{\bm x \in \mathcal{V}} \phi_{\bm x} \, ,
\end{align*}
and $\langle \cdot \rangle_{V, c}$ is the expectation value with respect to the partition function \eqref{eq:partition-function} for a system in a volume~$V$ in the presence of an external field~$c$.
The $Z(2)$ symmetry is said to be spontaneously broken, if  $\langle M \rangle \neq 0$ for a given fixed lattice spacing. 

Whether the $Z(2)$ symmetry can be spontaneously broken in the ground state at all depends on the number of spacetime dimensions. 
To be specific, the Mermin-Wagner theorem forbids \gls{SSB} in $d < 2$ spacetime dimensions which results in a vanishing magnetization, i.e., $\langle M \rangle = 0$ for $d < 2$, regardless of the exact values of the model parameters. 
Note that, for theories with a continuous symmetry, such as $O(N > 1)$, there is no \gls{SSB} even in $d = 2$ spacetime dimensions due to the presence of massless Nambu-Goldstone bosons. 

We emphasize that the role of the explicit symmetry breaking term in the definition of the magnetization \eqref{eq:magnetization} is crucial, since it distinguishes a direction in field space along which the formation of a non-trivial minimum is energetically favored, 
such that $\langle M \rangle_{V, c>0} > 0$.
Without external field~$c$, the magnetization would vanish for all finite volumes, i.e., $\langle M \rangle_{V, c=0} = 0$, and consequently, the limit in \cref{eq:magnetization} would vanish for all bare actions with a global~$Z(2)$ symmetry, regardless of the number of spacetime dimensions.

Quantum fluctuations associated with bosonic degrees of freedom tend to restore the symmetry in the ground state. 
Therefore, it is necessary (but not sufficient) to choose $m^2<0$ in Eq.~\eqref{eq:action-on-lattice} in order to obtain a ground state in the full quantum theory which is governed by spontaneous $Z(2)$ symmetry breaking. 
Indeed, provided~$m^2$ has been chosen smaller than a critical value which depends on the parameters~$d$ and~$\lambda$, the magnetization remains finite in $d \geq 2$ spacetime dimensions, even after all quantum fluctuations have been integrated out. 

\subsection{Effective potential}\label{sec:effective-potential}
Many physical observables of our model can be directly extracted from the effective potential $U$, which is the potential contribution of the effective action for vanishing external fields, $c=0$.
The effective potential inherits the $Z(2)$ symmetry of the bare action and is given by the Legendre transform of the Schwinger functional $W = \ln \mathcal{Z}$ at $c=0$ evaluated at a constant field configuration 
$\phi = (\phi_{\bm x})_{\bm x \in \mathcal{V}} = (\varphi, \, \dots, \varphi):$\footnote{The quantity $\phi$ should not be confused with a field vector as encountered in $O(N)$ models. The entries~$\varphi$ of this tuple are associated with the spacetime points~$\bm x$ and assume the same value at all spacetime points in case of a constant field configuration.}
\begin{align*}
	\numberthis\label{eq:effective-potential}
	U(\varphi) = \frac{1}{V} \sup_{J}\Big(\phi \cdot J - W_{c=0}[J]\Big) \, .
\end{align*}
Assuming the field configuration at the supremum, $J_\text{sup}$, is homogeneous, $J_\text{sup} = (j, \ldots, j)$,\footnote{%
	This assumption is in general only true when the condensate is homogeneous for all $c$, i.e., when $\langle \phi_{\bm x} \rangle_{V, c} = \varphi$ for all $\bm x \in \mathcal{V}$. Then we find
	\begin{align*}
		\varphi = \langle \phi_{\bm x} \rangle_{V, c} = a^{-d}\frac{\partial W}{\partial J_{\bm x}} \Big|_{J = (c, \dots, c)} \, .
	\end{align*}
	In particular, this holds when translation invariance is preserved.} 
and noting that the external field~$c$ enters the partition function in the same way as a source, \cref{eq:effective-potential} can be rewritten as follows:
\begin{align*}
	\numberthis\label{eq:effective-potential-via-explicit-symmetry-breaking}
	U(\varphi) &= \frac{1}{V} \sup_{c}\Big(V \varphi \, c - W_{c}[0]\Big) \, .
\end{align*}

Furthermore, since the magnetization can be expressed as a derivative of the Schwinger functional with respect to the external field,
\begin{align*}
	\numberthis\label{eq:M-via-schwinger-functional}
	\langle M \rangle_{V, c} = \frac{1}{V}\frac{\partial}{\partial c} W \, ,
\end{align*}
we conclude, together with \cref{eq:effective-potential-via-explicit-symmetry-breaking}, that
\begin{align*}
	\numberthis\label{eq:quantity-phi0}
	\partial_\varphi U(\langle M \rangle_{V, c}) = c \, .
\end{align*}

From \cref{eq:quantity-phi0}, we can already infer some general properties of the $Z(2)$-symmetric effective potential in both finite and infinite volume.
In finite volume the effective potential is strictly convex with a trivial global minimum at $\varphi = 0$, since $\langle M \rangle_{V, c} \to 0$ as $c \to 0$.
Only in the thermodynamic limit, $V\to \infty$, when a non-trivial magnetization persists as $c\to 0$, the effective potential $U$ has two degenerate non-trivial minima located at $\pm \varphi_0$ which must also coincide with the magnetization~\eqref{eq:magnetization}, i.e., $\varphi_0 \equiv \langle M \rangle$.

Moreover, \cref{eq:quantity-phi0} can be used to reconstruct the effective potential and its derivative, $\partial_{\varphi} U(\varphi)$, by performing multiple calculations of the magnetization for different values of $c$,  see also Refs.~\cite{Gao:2021vsf,Braun:2023qak}.
This approach is exactly what we employ in our analysis of finite systems in \cref{sec:comparison}.

Another physically relevant quantity which we will discuss in \cref{sec:comparison} is the so-called susceptibility, which is the integrated connected two-point correlation function and can be expressed by the magnetization:
\begin{align*}
	\numberthis\label{eq:chi}
	\chi_{V, c} &= V \langle (M - \langle M \rangle_{V, c})^2 \rangle_{V, c} \, .
\end{align*}
This quantity diverges at second-order phase transitions in the thermodynamic limit and hence can be used to identify these.
In terms of the Schwinger functional, it can be written as the second derivative with respect to the external field $c$,
\begin{align*}
	\numberthis\label{eq:chi-via-schwinger-functional}
	\chi_{V, c} = \frac{1}{V}\frac{\partial^2}{\partial c^2} W \, ,
\end{align*}
which, using \cref{eq:M-via-schwinger-functional,eq:quantity-phi0}, implies 
\begin{align*}
	\numberthis\label{eq:quantity-chi}
	\partial^2_\varphi U(\langle M \rangle_{V, c}) = \chi_{V, c}^{-1} \, .
\end{align*}
Thus, the susceptibility is associated to the inverse curvature of the effective potential evaluated at the corresponding magnetization $\varphi = \langle M \rangle_{V, c}$.
In finite volumes, the susceptibility can never diverge as the effective potential is strictly convex, regardless of the number of spacetime dimensions or the specific values of $m^2$ and $\lambda > 0$ in the bare potential.

\section{Lattice \gls{MC} Simulations}\label{sec:MC-simulation-intro}
The aim of this work is a direct comparison of lattice \gls{MC} and lattice \gls{FRG} calculations. 
By using the same discretized action on the same spacetime lattice with given lattice spacing and volume, we avoid 
any ``translation'' or renormalization of model parameters between the two approaches, and in particular the necessity of a continuum limit. 

For given lattice spacing, volume and bare parameter sets, the only approximation of a \GLS{MC} simulation consists of evaluating 
the path integral on a finite (rather than infinite) number of field configurations. In a process referred to as importance sampling, 
a set of field configurations is generated with probability weights given by the Boltzmann factor $e^{-S[\phi]}$.  The expectation value of a given observable~$\mathcal O$ is then approximated as an average over the generated field configurations, 
\begin{align*}
	\numberthis\label{eq:partition-function-as-sum-over-configurations}
	\langle \mathcal{O} \rangle_{V, c} \approx \frac{1}{N} \sum_{i=1}^{N} \mathcal{O}[\phi^i] \,.
\end{align*}
Here, $\phi^i$ refers to a specific field configuration generated in the \gls{MC} process. The fluctuation of the observable
with the different configurations is evaluated by the usual standard deviation, which diminishes as $N^{-1/2}$ as the number of
configurations is increased.  
 
We generate our field configurations using a \gls{HMC} algorithm~\cite{DUANE1987216}. 
Furthermore, after every \gls{HMC} step we include a sign flip $\phi \rightarrow -\phi$ with a subsequent accept-reject step. 
This ensures that the simulation does not get ``stuck" in a specific minimum of the potential, thus reducing the initial correlation between
consecutive configurations, and therefore the overall simulation time. 
In order to control and further suppress correlations, we bin our data and calculate the statistical errors using the jackknife procedure. 
Since a scalar theory on the lattice is computationally not very demanding, the statistical uncertainty in the following results could be kept small by accumulating a large amount of uncorrelated data. Whenever not visible, error bars are smaller than the symbol sizes.

\section{Lattice Functional Renormalization Group}\label{sec:lfrg}
Although the \gls{FRG} method has originally been developed for studies of systems in infinitely large, continuous spacetime volumes, it is also suitable to study theories on finite spacetime lattices. 
This has been done in previous works on scalar field theories such as in Refs.~\cite{Dupuis:2008vr,Machado:2010wi,Rancon:2011aa,Krieg:2017hcm,Pawlowski:2017rhn}. 
Studies of systems of scalar field theories and fermion-boson models in a continuous but finite spacetime volume have been put forward in Refs.~\cite{Braun:2004yk,Braun:2005gy,Braun:2009ruy,Fister:2015eca} which have been supplemented with an analysis of finite-temperature and density effects~\cite{Braun:2010vd,Braun:2011iz,Tripolt:2013zfa}, see Ref.~\cite{Klein:2017shl} for a review. 

The underlying idea of the \gls{FRG} approach is to integrate out the momentum modes of the partition function successively, starting with the bare action~\eqref{eq:action-on-lattice}. 
To this end, it is necessary to introduce an infrared regulator $R_k$ which introduces a \gls{RG} scale $k$ into the theory. 
This regulator suppresses modes with momenta $\epsilon_{\bm q} \lesssim k$ while modes with momenta $\epsilon_{\bm q} \gtrsim k$ are no longer affected by the regulator. 
In the path integral, the regulator appears in form of a regulator term,
\begin{align*}
	\numberthis\label{eq:regulator-term}
	\Delta S_k(\{\phi_{\bm x \in \mathcal{V}}\}) =\frac{1}{V}\sum_{\bm q \in \tilde{\mathcal{V}}} \frac{R_{k}(\epsilon_{\bm q})}{2}\tilde\phi_{-\bm q}  \tilde\phi_{\bm q} \, ,
\end{align*}
which is added to the bare action $S$, see Refs.~\cite{Wetterich:1992yh,Pawlowski:2005xe} for a general discussion of the properties of regulators.
This yields the scale-dependent partition function~$\mathcal{Z}_k$,  
\begin{align*}
	\numberthis\label{eq:scale-dependent-partition-function}
	\mathcal{Z}_k[J] =  \int \mathcal{D} \phi \, e^{-S[\phi] - \Delta S_k[\phi] + J \cdot \phi} \, ,
\end{align*}
from which the Wetterich equation can be derived in a similar way as for continuous spacetimes. 
The exact flow equation for the effective action in position space reads~\cite{Wetterich:1992yh}
\begin{align*}
	\numberthis\label{eq:wetterich-equation-on-lattice}
	\partial_t \bar\Gamma_k[\phi] = a^{2d}\sum_{\bm x, \bm y \in \mathcal{V}} \Big(\bar\Gamma^{(2)}_{k, \bm a\bm b}[\phi] + \Delta S^{(2)}_{k,\bm a \bm b}\Big)^{-1}_{\bm x \bm y} \partial_t \Delta S^{(2)}_{k,\bm y \bm x} \,.
\end{align*}
Here, we introduced the \gls{RG} scale derivative $\partial_t = - k \partial_k$,\footnote{
	We define the functional derivatives of the $n$-th order acting on an action $A$ on a spacetime lattice in position space as
\begin{align*}
	A^{(n),\, \bm x_1 \dots \bm x_n}[J] = a^{-d}\frac{\partial}{\partial J_{\bm x_1}} \dots a^{-d}\frac{\partial}{\partial J_{\bm x_n}} A[J] \, ,
\end{align*}
and correspondingly in momentum space as
\begin{align*}
	A^{(n),\, \bm q_1 \dots \bm q_n}[J] = V\frac{\partial}{\partial \tilde J_{\bm q_1}} \dots V\frac{\partial}{\partial \tilde J_{\bm q_n}} A[J] \, .
\end{align*}
}
where $t$ can be related to the so-called \gls{RG} time. 
For reviews and introductions to the continuum formulation of the Wetterich equation, see Refs.~\cite{Berges:2000ew,Pawlowski:2005xe,Gies:2006wv,Delamotte:2007pf,Kopietz:2010zz,Braun:2011pp,Dupuis:2020fhh}.

The lattice formulation of the Wetterich equation~\eqref{eq:wetterich-equation-on-lattice} is a partial differential equation with $1 + \vert \mathcal{V}\vert$ variables. 
Its solution, the scale-dependent effective average action $\bar\Gamma_k$, interpolates between the bare action $S[\phi]$ as $k \to \infty$ and the full quantum effective action $\Gamma[\phi]$ as $k \to 0$. The latter property of the scale-dependent effective action is trivially fulfilled, as \cref{eq:scale-dependent-partition-function} reduces to \cref{eq:partition-function} when $k \to 0$. A more detailed derivation of the \gls{UV} limit is shown in \cref{sec:uv-regime}. In the following, we refer to the limits $k \to \infty$ and $k \to 0$ as \gls{UV} limit and \gls{IR} limit, respectively.

Formally, the Wetterich equation represents an initial value problem where the initial condition is given by an action $\bar\Gamma_{\Lambda}[\phi]$ at the so-called cutoff scale $\Lambda$ and the differential equation is given by the Wetterich equation \eqref{eq:wetterich-equation-on-lattice}. 
As long as $\Lambda$ is finite, this action is not identical to the bare action $S$. 
However, in the \gls{UV} limit, the ``running couplings'' of the \gls{FRG} flow, i.e., the couplings $\lambda_i(k)$ of $\bar\Gamma_k$, must approach the (finite) values of the corresponding couplings $\lambda_i$ in the bare action~$S$:
\begin{align*}
	\numberthis\label{eq:uv-limit}
	\lim_{k \to \infty} \lambda_i(k) = \lambda_i \, .
\end{align*}
Consequently for large \gls{RG} scales $k \gg 1/a$, we should observe that the change of the couplings with respect to the \gls{RG} scale approaches zero, i.e.,
\begin{align*}
	\numberthis\label{eq:uv-behavior}
	\partial_t \lambda_i(k) \approx 0 \quad \text{for $k \gg 1/a$} \, .
\end{align*}
To obtain the full quantum effective action $\Gamma$, it is therefore sufficient to initialize the flow equation \eqref{eq:wetterich-equation-on-lattice} with the action $\bar\Gamma_{\Lambda} = S$ at some large but finite cutoff. 
This also ensures that this initial condition {\it canonically} fulfills the requirement of \gls{RG} consistency~\cite{Braun:2018svj}, i.e., the cutoff-independence of the full quantum effective action.
This is in contrast to continuum theories, where a non-trivial scale-dependent initial condition $S_\Lambda$ must be determined to ensure that the full quantum effective action $\Gamma$ remains unchanged as $\Lambda$ is varied.

\subsection{\gls{RG} flow on finite spacetime lattices}

In this subsection, we discuss several aspects of \gls{RG} flows on finite spacetime lattices. 

Depending on the dispersion relation, one obtains a finite set of kinetic energy levels, $\mathcal{E} = \{\epsilon_{\bm q} \,|\, \bm q \in \tilde{\mathcal{V}}\} = \{0, \Delta \epsilon, \dots, \epsilon_{\mathrm{max}}\}$. 
Here, $\Delta \epsilon$ is the lowest non-zero kinetic energy level, i.e., $\Delta \epsilon = \min_{\bm q \in \tilde{\mathcal{V}} \setminus {\bm 0}}(\epsilon_{\bm q})$. 
For example, for the relation \eqref{eq:dispersion-relation}, the highest kinetic energy level is~$\epsilon_{\mathrm{max}} = 2\sqrt{d}/a$ and the lowest non-zero level is given by $\Delta \epsilon = 2 \sin(\pi/N_{\sigma})/a$.
This allows us to divide the \gls{RG} flow into three regimes: the \gls{UV} regime with $k > \epsilon_{\mathrm{max}}$, the intermediate regime with $\Delta \epsilon \leq k \leq \epsilon_{\mathrm{max}}$ and the \gls{IR} regime with $k < \Delta\epsilon$.
It is important to note that the precise values of the boundaries of these regimes may shift when the regulator is changed. This is because the regulator itself defines the notion of the \gls{RG} scale. However, for the Litim regulator~\cite{Litim:2001up,Litim:2001fd},
\begin{align*}
	\numberthis\label{eq:litim-regulator}
	R_k(\epsilon_{\bm q}) &= (k^2 - \epsilon^2_{\bm q}) \Theta(k^2 - \epsilon^2_{\bm q}) \, ,
\end{align*}
which we shall primarily use in this work, the boundaries of the different regimes are as defined above.

Before discussing the different regimes of the \gls{RG} flow, we introduce useful definitions and relations which will help in analyzing the dynamics in these regimes below. 
We start by considering the Wetterich equation~\eqref{eq:wetterich-equation-on-lattice} in momentum space and exploit the fact that the regulator is diagonal in momentum space, cf. \cref{eq:regulator-term}:
\begin{align*}
	\numberthis\label{eq:wetterich-equation-energy-specturm-decompositon}
	\partial_t \bar\Gamma_k[\phi] 
	&= \frac{1}{2} \frac{1}{V}\sum_{\bm q \in \tilde{\mathcal{V}}} G^{(2)}_{k,\bm q, -\bm q}[\phi] \, \partial_t R_k(\epsilon_{\bm q}) \, ,
\end{align*}
with the propagator
\begin{align*}
	\numberthis\label{eq:general-momentum-propagator}
	G^{(2)}_{k,\bm p,\bm q}[\phi] = \Big(\bar\Gamma^{(2)}_{k, \bm a\bm b}[\phi] + \Delta S^{(2)}_{k,\bm a \bm b}\Big)^{-1}_{\bm p \bm q} \, .
\end{align*}
In general, the inversion of the regularized two-point function in momentum space is non-trivial, even on a finite spacetime lattice. 
However, employing that the system under consideration is translation invariant and evaluating the propagator \eqref{eq:general-momentum-propagator} at a constant background field configuration, $\phi = (\phi_{\bm x})_{\bm x \in \mathcal{V}} = (\varphi,\dots,\varphi)$, we have
\begin{align*}
	\numberthis\label{eq:general-momentum-propagator-evaluated-at-background-field}
	G^{(2)}_{k,\bm p, \bm q}[\phi] = G^{(2)}_k(\varphi, {\bm q}) \, V \delta_{\bm p, -\bm q} \, 
\end{align*}
and therefore
\begin{align*}
	\numberthis\label{eq:general-momentum-propagator-2}
	G^{(2)}_{k}(\varphi, {\bm q}) = \frac{1}{\partial^2_\varphi U_k(\varphi) + \Delta \bar\Gamma^{(2)}_k(\varphi, {\bm q}) + R_k(\epsilon_{\bm q})} \, .
\end{align*}
Here, we have divided $\bar\Gamma^{(2)}_{k,\bm p, \bm q}[\phi] = \bar\Gamma^{(2)}_k(\varphi, {\bm q}) \, V \delta_{\bm p, -\bm q}$ into a potential-like and a kinetic-like contribution. For the potential-like contribution we have 
\begin{align*}
	\numberthis\label{eq:propagator-potential-contribution}
	\partial^2_\varphi U_k(\varphi) = {\bar\Gamma}^{(2)}_k(\varphi, {\bm 0}) \, ,
\end{align*}
where $U_k$ corresponds to the potential term of the scale-dependent effective average action $\bar\Gamma_k$, i.e., $U_k(\varphi) = V^{-1} \bar\Gamma_k[\phi]$. All momentum-dependent terms are encoded in the kinetic-like contribution:
\begin{align*}
	\numberthis\label{eq:propagator-kinetic-contribution}
	\Delta{\bar\Gamma}^{(2)}_k(\varphi, {\bm q}) = {\bar\Gamma}^{(2)}_k(\varphi, {\bm q}) - {\bar\Gamma}^{(2)}_k(\varphi, {\bm 0}) \, .
\end{align*}
\subsubsection{Infrared regime}\label{sec:ir-regime} 
Using the Litim regulator \eqref{eq:litim-regulator} or any other regulator fulfilling the property
\begin{align*}
	\numberthis\label{eq:ir-regulator-property}
	\partial_t R_k(\epsilon_{\bm q}) = 0 \quad \text{for $k \leq \epsilon_{\bm q}$} \, ,
\end{align*}
the Wetterich equation \eqref{eq:wetterich-equation-energy-specturm-decompositon} yields 
\begin{align*}
	\numberthis\label{eq:wetterich-equation-energy-specturm-decompositon-2}
	\partial_t \bar\Gamma_k[\phi] = \frac{1}{2}\frac{1}{V}\sum_{\substack{\bm q \in \tilde{\mathcal{V}} \\ \epsilon_{\bm q} < k}} G^{(2)}_{k,\bm q, -\bm q}[\phi] \, \partial_t R_{k}(\epsilon_{\bm q}) \, .
\end{align*}
This choice of regulator canonically truncates the right hand side of the Wetterich equation without assuming any approximation. 
This implies, that in the \gls{IR} regime, for $k < \Delta\epsilon$, only the zero-mode contributes to the Wetterich equation, which reduces \cref{eq:wetterich-equation-energy-specturm-decompositon-2} to 
\begin{align*}
	\numberthis\label{eq:wetterich-equation-in-ir-regime}
	\partial_t \bar\Gamma_k[\phi] = \frac{1}{2} \frac{1}{V}G^{(2)}_{k,\bm 0, \bm 0}[\phi] \, \partial_t R_{k}(0) \, .
\end{align*}

Now, evaluating both sides of \cref{eq:wetterich-equation-in-ir-regime} at a constant background field configuration $\phi = (\varphi, \dots, \varphi)$ and using the structure of the propagator \eqref{eq:general-momentum-propagator-2} with $\Delta{\bar\Gamma}^{(2)}_k(\varphi, {\bm 0}) = 0$, we find for the scale-dependent effective potential,  $U_k(\varphi) = V^{-1} \bar\Gamma_k[\phi]$, the exact flow equation
\begin{align*}
	\numberthis\label{eq:effective-potential-ir-regime}
	\partial_t U_k(\varphi) &= \frac{1}{V}\frac{\partial_t R_{k}(0)}{\partial^2_\varphi U_k(\varphi) + R_{k}(0)} \, .
\end{align*}
This flow equation yields a well-defined solution as $k \to 0$, as long as the regulator is masslike~\cite{Litim:2001up,Litim:2001fd}, i.e., $R_{k > 0}(0) > 0$.
Furthermore, it functionally mimics a zero-dimensional \gls{RG} flow, which must lead to a strictly convex quantum effective action in the \gls{IR} limit, see Refs.~\cite{Keitel:2011pn,Koenigstein:2021syz,Koenigstein:2021rxj,Steil:2021cbu,Zorbach:2024rre}.
This is in accordance with the general absence of spontaneous symmetry breaking on finite spacetime lattices.

Note that due to the evaluation of the propagator \eqref{eq:general-momentum-propagator-2} at $\bm q = \bm 0$, the flow equation for the scale-dependent effective potential completely decouples from the non-trivial kinetic structure $\Delta\bar\Gamma^{(2)}_k$. Meaning that, in the \gls{IR} regime, the scale-dependent effective potential is not affected by couplings like a wave-function renormalization or other couplings associated to the momentum structure of the scale-dependent effective action.

Finally, note that the assumption of translational invariance, used in \cref{eq:general-momentum-propagator-evaluated-at-background-field}, is almost always inherent in the truncation ansatz employed in \gls{FRG} studies.

\subsubsection{Ultraviolet regime}\label{sec:uv-regime}

Let us now discuss the \gls{UV} regime of the \gls{RG} flow. 
Specifically, we focus on this regime for regulators to which we refer as lattice site decoupling regulators. 
These are regulators which, above a certain \gls{RG} scale~$k^{\star}$ -- the lattice site decoupling scale -- eliminate any kinetic structure in the propagator. 
As a result, the scale-dependent partition function $\mathcal{Z}_k[J]$ in \cref{eq:scale-dependent-partition-function} reduces to a product of zero-dimensional partition functions $\mathcal{Z}_k^{0d}(J_{\bm x})$. 
Meaning that the regulator term in the action renders all fluctuations purely local in this regime, see Ref.~\cite{Machado:2010wi}. 

More precisely, in order to qualify as lattice site decoupling, the regulator must fulfill 
\begin{align*}
	\numberthis\label{eq:lattice-site-decoupling-regulator-property}
	R_{k > k^{\star}}(\epsilon_{\bm q}) = M^2_k - \epsilon^2_{\bm q} \, ,
\end{align*}
for all \gls{RG} scales $k$ greater than the decoupling scale~$k^\star$, which implies that
\begin{align*}
	\numberthis\label{eq:action-plus-regulator-term}
	S[\phi] + \Delta S_{k > k^{\star}}[\phi] &= a^d \sum_{\bm x \in \mathcal{V}} \frac{M^2_k}{2}  \phi_{\bm x}^2 + a^d \sum_{\bm x \in \mathcal{V}} \mathcal{U}(\phi_{\bm x})
\end{align*}
is a purely local action. 
Here, $M_k$ is a $k$-dependent mass term to be chosen such that it diverges as $k \to \infty$.
This property guarantees that the scale-dependent effective average action $\bar\Gamma_k$ approaches the bare action $S$ in the \gls{UV} limit, as we shall show below.
For example, the Litim regulator \eqref{eq:litim-regulator} with $k^\star = \epsilon_{\mathrm{max}}$ and $M_k = k$ is of this type.\footnote{
	Another example for a lattice site decoupling regulator is the smooth Litim regulator introduced in Ref.~\cite{Zorbach:2024zjx} which has $k^\star = \sqrt{2} \epsilon_{\mathrm{max}}$ as its lattice site decoupling scale.
}
For the scale-dependent partition function \eqref{eq:scale-dependent-partition-function} we find
\begin{align*}
	\numberthis\label{eq:partition-function-lattice-site-decoupling}
	\mathcal{Z}_{k > k^{\star}}[J] &= \int \mathcal{D} \phi \, \prod_{\bm x \in \mathcal{V}} e^{-a^d \left(\frac{1}{2}M^2_k \phi_{\bm x}^2 + \mathcal{U}(\phi_{\bm x}) - J_{\bm x} \phi_{\bm x}\right)} \\
	&= \prod_{\bm x \in \mathcal{V}} \mathcal{Z}^{0d}_k(a^d J_{\bm x}) \, , 
\end{align*}
where we have used \cref{eq:path-integral-measure} and introduced the zero-dimensional partition function
\begin{align*}
	\numberthis\label{eq:0d-partition-function}
	\mathcal{Z}^{0d}_k(j) = a^{d_{\phi}} \int^{\infty}_{-\infty} \d \varphi \, e^{-S_k^{0d}(\varphi) + j \varphi} \, ,
\end{align*}
with $S_k^{0d}(\varphi) = a^d (\frac{1}{2}M^2_k \varphi^2 + \mathcal{U}(\varphi))$.
For the Schwinger functional we analogously find
\begin{align*}
	\numberthis\label{eq:schwinger-functional-lattice-site-decoupling}
	W_{k > k^{\star}}[J] 
	&= \sum_{\bm x \in \mathcal{V}} W^{0d}_{k}(a^d J_{\bm x}) \, ,
\end{align*}
where $W^{0d}_{k}(j) = \ln\mathcal{Z}^{0d}_k(j)$.
Due to the simple structure of \cref{eq:schwinger-functional-lattice-site-decoupling}, we find
\begin{align*}
	\numberthis\label{eq:schwinger-functional-lattice-site-decoupling-derivative}
	a^{-d}\frac{\partial}{\partial J_{\bm x}} W_{k > k^{\star}}[J] = W^{0d\,(1)}_k(a^d J_{\bm x}) \,.
\end{align*}
This implies that $J_{\bm x}[\phi] = a^{-d} [W^{0d\,(1)}_k]^{-1}(\phi_{\bm x})$. 
Hence, the Legendre transform of $W_k[J]$, i.e., the scale-dependent effective action, reads
\begin{align*}
	\numberthis\label{eq:scale-dependent-effective-action-lattice-site-decoupling-derivative}
	\Gamma_{k > k^{\star}}[\phi] &= a^d \sum_{\bm x \in \mathcal{V}}J_{\bm x}[\phi] \phi_{\bm x} - W_{k > k^{\star}}[J[\phi]] \\
	&= \sum_{\bm x \in \mathcal{V}}\left(
	[W^{0d\,(1)}_k]^{-1}(\phi_{\bm x}) \phi_{\bm x}  \right. \\ 
	& \qquad\qquad  \left. - W^{0d}_{k}([W^{0d\,(1)}_k]^{-1}(\phi_{\bm x}))
	\right) \\
	&= \sum_{\bm x \in \mathcal{V}} \Gamma^{0d}_{k}(\phi_{\bm x}) \, .
\end{align*}
The modified Legendre transform, the scale-dependent effective average action, is then given by
\begin{align*}
	\numberthis\label{eq:scale-dependent-effective-average-action-lattice-site-decoupling-derivative}
	\bar \Gamma_{k > k^{\star}}[\phi] &= \sum_{\bm x \in \mathcal{V}} \Gamma^{0d}_{k}(\phi_{\bm x}) - \Delta S_k[\phi] \\
	&= \frac{1}{V}\sum_{\bm q \in \tilde{\mathcal{V}}} \frac{\epsilon^2_{\bm q}}{2}\tilde\phi_{-\bm q}  \tilde\phi_{\bm q} \\
	& \qquad + \sum_{\bm x \in \mathcal{V}} \left( \Gamma^{0d}_{k}(\phi_{\bm x}) - a^d \frac{M^2_k}{2} \phi^2_{\bm x} \right)\, .
\end{align*}	
It is worth mentioning that, due to the regulator term~$\Delta S_k[\phi]$ in the modified Legendre transformation in \cref{eq:scale-dependent-effective-average-action-lattice-site-decoupling-derivative}, a kinetic contribution is added to the scale-dependent effective average action. 
Furthermore, since we did not yet make any approximation, the solution \eqref{eq:scale-dependent-effective-average-action-lattice-site-decoupling-derivative} is exact which means that it is a solution of the (untruncated) Wetterich equation~\eqref{eq:wetterich-equation-on-lattice} for $k > k^\star$. 
Only the term associated with the potential in $\bar\Gamma_k$ changes during the \gls{RG} flow, other couplings are {\it not} generated. 
This reflects the purely local structure of the theory for \gls{RG} scales $k > k^{\star}$. 
From this analysis, we conclude that the use of a lattice site decoupling regulator is advisable in actual applications of our lattice \gls{FRG} framework.

Finally, using \cref{eq:scale-dependent-effective-average-action-lattice-site-decoupling-derivative}, we can prove that the scale-dependent effective average action indeed approaches the bare action $S$ as $k\to\infty$. 
To this end, we note that $[W^{0d\,(1)}_k]^{-1}(\phi_{\bm x}) \sim a^d M^2_k \phi_{\bm x}$ as $k\to \infty$ which implies
\begin{align*}
	\numberthis\label{eq:uv-limit-proof}
	&\Gamma^{0d}_k(\phi_{\bm x}) - a^d \frac{M^2_k}{2} \phi_{\bm x}^2 \\
	&\quad= - \ln( a^{d_{\phi}} \int^{\infty}_{-\infty} \d \varphi \, e^{-a^d \mathcal{U}(\varphi)} \,  e^{-a^d \frac{M^2_k}{2}\left( \varphi - \phi_{\bm x} \right)^2}) \\
	&\quad\sim a^d \mathcal{U}(\phi_{\bm x}) + C 
\end{align*}
for~$k\to \infty$. 
Here, $C$ is a field-independent and thus irrelevant constant.
We would like to stress that each term in \cref{eq:uv-limit-proof} diverges separately. 
However, the combination of all terms yields the finite bare potential in the \gls{UV} limit and thus $\bar\Gamma_k[\phi] \to S[\phi]$ as $k\to\infty$. 
Note that this also implies the \gls{UV} behavior of the couplings as shown  in \cref{eq:uv-behavior}.

\subsubsection{Intermediate regime}\label{sec:intermediate-regime}
In the intermediate regime, $\Delta\epsilon \leq k \leq \epsilon_{\mathrm{max}}$, the \gls{RG} flow is non-trivial and in principle all couplings allowed by the symmetries of a given model are dynamically generated. 
We add that the size of this regime shrinks as we decrease the number of lattice sites~$N_{\sigma}$ and disappears for~$N_{\sigma}=1$.

\subsection{Local potential approximation}

As we have already seen in \cref{sec:ir-regime} and \cref{sec:uv-regime}, the effective potential in the scale-dependent effective average action plays a dominant role in the \gls{IR} as well as in the \gls{UV} regime, especially in case of regulators satisfying the properties \eqref{eq:ir-regulator-property} and \eqref{eq:lattice-site-decoupling-regulator-property}.
To describe these regimes accurately, it is therefore mandatory to consider an approximation of the Wetterich equation \eqref{eq:wetterich-equation-on-lattice} which is exact in these limits. 
This is the case for the \gls{LPA} as these limits correspond to zero-dimensional theories. 
In the following we shall therefore employ this approximation which represents the lowest order of the derivative expansion but already goes beyond the mean-field approximation as it takes into account fluctuation effects. 

The \gls{LPA} assumes in every \gls{RG} step the ansatz 
\begin{align*}
	\numberthis\label{eq:LPA-ansatz}
	\bar\Gamma_k[\phi] = \frac{1}{V} \sum_{\bm q \in \tilde{\mathcal{V}}} \frac{\epsilon^2_{\bm q}}{2} \tilde\phi_{-\bm q} \tilde\phi_{\bm q} + a^d \sum_{\bm x \in {\mathcal{V}}} U_k(\phi_{\bm x}) 
\end{align*}
for the scale-dependent effective average action on the right-hand side of the Wetterich equation \eqref{eq:wetterich-equation-on-lattice}.
This implies that terms associated with derivatives of the fields enter the right-hand side of the Wetterich equation only in the form as they appear in the classical action.
Nevertheless, within \gls{LPA}, such couplings are dynamically generated, especially in the aforementioned intermediate regime of the \gls{RG} flow, and can in principle be straightforwardly calculated by taking field derivatives on both sides of the Wetterich equation~\eqref{eq:wetterich-equation-energy-specturm-decompositon}.

Using \cref{eq:LPA-ansatz} as truncation for the scale-dependent effective action, the kinetic contribution is simply given by $\Delta \bar\Gamma^{(2)}_k(\varphi, {\bm q}) = \epsilon^2_{\bm q}$, and thus the propagator \eqref{eq:general-momentum-propagator-2} reads
\begin{align*}
	\numberthis\label{eq:propagator-in-LPA}
	G^{(2)}_{k}(\varphi, {\bm q}) = \frac{1}{\partial^2_\varphi U_k(\varphi) + \epsilon^2_{\bm q} + R_k(\epsilon_{\bm q})} \, .
\end{align*}
In particular, for the Litim regulator \eqref{eq:litim-regulator}, we have $\epsilon^2_{\bm q} + R_k(\epsilon_{\bm q}) = k^2$ for $k \leq \epsilon_{\bm q}$.
Hence, evaluating the Wetterich equation \eqref{eq:wetterich-equation-energy-specturm-decompositon} at a constant background field configuration, $\phi = (\varphi,\dots,\varphi)$, and using the Litim regulator, the flow equation for the scale-dependent effective potential reads
\begin{align*}
	\numberthis\label{eq:LPA-flow-equation}
	\partial_t U_k(\varphi) &= \frac{1}{2} \Omega(k) \frac{\partial_t k^2}{k^2 + \partial^2_{\varphi}U_k (\varphi)} \, ,
\end{align*}
where $\Omega(k)$ is the density of modes:
\begin{align*}
	\numberthis\label{eq:mode-density}
	\Omega(k) = \frac{1}{V}\sum_{\bm q \in \tilde{\mathcal{V}}} \Theta(k^2 - \epsilon^2_{\bm q}) \, .
\end{align*}
In the \gls{UV} regime, $k > \epsilon_{\mathrm{max}}$, we find $\Omega(\epsilon) = a^{-d}$.
In the \gls{IR} regime, $k < \Delta\epsilon$, we have $\Omega(k) = 1/V$\footnote{
	For comparison, in a continuous and infinite spacetime, we find
	\begin{align*}
		\Omega(k) = \frac{\mathrm{surf}(d)}{(2\pi)^d} \frac{1}{d} k^{d} \, ,
	\end{align*}
	where $\mathrm{surf}(d)$ is the surface of a $d$-dimensional unit sphere.
}.
We emphasize that, even in this approximation, the flow equation~\eqref{eq:LPA-flow-equation} already represents a highly nonlinear diffusion equation. 
In \cref{sec:solving-the-flow-equation}, we discuss the numerical setup used to solve this differential equation in the present work. 

\begin{figure}
	\includegraphics{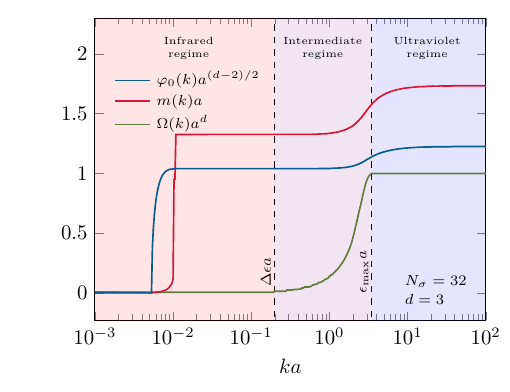}
	\caption{\label{fig:regimes}Illustration of the \gls{RG} flow of the (global) minimum~$\varphi_0(k)$ of the potential~$U_k$, the curvature mass~$m(k)$ evaluated at the minimum~$\varphi_0(k)$ of~$U_k$, and the density of modes~$\Omega(k)$
	in the \gls{IR}, intermediate and \gls{UV} regimes for a given bare action, see main text for details.}
\end{figure}

In \cref{fig:regimes}, we illustrate the behavior of various quantities in the \gls{RG} flow for the bare action \eqref{eq:action-on-lattice} in $d=3$ dimensions with $a\lambda = 6$ and $(am)^2 = -1$.
To be specific, we show the scale-dependent (global) minimum $\varphi_0(k)$ of the scale-dependent potential $U_k$, the curvature mass $m(k)$ evaluated at $\varphi_0(k)$ of~$U_k$, and the density of modes $\Omega(k)$ as functions of the \gls{RG} scale $k$. 
We observe that $\varphi_0(k)$ and $m(k)$ approach plateaus, reflecting the convergence of $U_k \to \mathcal{U}$ in the \gls{UV} limit, i.e., as $k\to \infty$.
From this figure we also deduce that, for a given bare action, it is indeed possible to find a finite initial \gls{RG} scale that is sufficiently large to suppress artefacts associated with its finiteness. 
For the specific parameter set represented in \cref{fig:regimes}, we find that $\Lambda = 100/a$ is sufficiently large.
We add that the \gls{RG} flow is exact down to the lattice site decoupling scale $k^\star = \epsilon_{\mathrm{max}}$, as discussed in \cref{sec:uv-regime}.
The density of modes~$\Omega(k)$ remains constant in this regime. 

In the intermediate regime, the \gls{RG} flow in \gls{LPA} is no longer exact. 
Here, the mode density decreases as the \gls{RG} scale is lowered until it reaches $\Omega(k = \Delta\epsilon) = 1/V$. 
The corresponding scale defines the onset of the \gls{IR} regime. 
In this regime, the flow equation for the scale-dependent effective potential completely decouples from kinetic contributions and only the zero mode contributes to the \gls{RG} flow, see \cref{eq:wetterich-equation-in-ir-regime}.
The \gls{LPA} is then no longer an approximation but exact again.
As a consequence of the fact that the \gls{RG} flow reduces to that of a zero-dimensional system in this regime, the minimum $\varphi_0(k)$ eventually approaches zero for $k \to 0$. 
Thus, there is no spontaneous symmetry breaking in the \gls{IR} limit, as it should be for zero-dimensional systems. 
The curvature mass $m(k)$ approaches a small positive value, indicating the formation of a very flat but strictly convex effective potential in the \gls{IR} limit.
Note that the flow equation associated with this regime, which can be extracted from \cref{eq:LPA-flow-equation} by replacing the mode density with~$1/V$, is indeed reminiscent of that of a zero-dimensional system, see also \cref{sec:ir-regime}.

\section{Results} 
\label{sec:res}
We begin the discussion of our numerical results by noting that we shall choose~$\lambda a^{4-d} = 6$ for the quartic coupling for all spacetime dimensions $d$ considered in this work. 
Thus, with respect to the parameters of our model, we vary only the squared bare mass parameter~$m^2$ and the external field~$c$, which is sufficient for a study of \gls{SSB} and phase transitions. 
All dimensionful quantities shall be given in units of the lattice spacing~$a$. 
For notational convenience, we are using natural units, ``$a=1$'', from here on.  

\subsection{FRG: Assessing LPA}\label{sec:assessing-lpa}
To obtain an intrinsic check of the reliability of \gls{LPA}, we analyze the kinetic term in the propagator \eqref{eq:general-momentum-propagator-2}. 
To that end, we derive the flow equation for the quantity $\Delta{\bar\Gamma}^{(2)}_k(\varphi, {\bm p})$ defined in \cref{eq:propagator-kinetic-contribution}.
This is done by first taking two field derivatives on both sides of the Wetterich equation \eqref{eq:wetterich-equation-energy-specturm-decompositon} and evaluating the resulting flow equation on a constant background field configuration $\phi = (\varphi, \dots, \varphi)$. 
From this, we then obtain the following expression in \gls{LPA}:
\begin{align*}
	\numberthis\label{eq:LPA-flow-equation-for-kinetic}
	\partial_t \Delta{\bar\Gamma}^{(2)}_k(\varphi, {\bm p})
	&= \frac{1}{V} \sum_{\bm q \in \tilde{\mathcal{V}}} \partial_t R_k(\epsilon_{\bm q})  \left(G^{(2)}_k(\varphi,{\bm q})  \, \partial^3_{\varphi}U_k(\varphi)\right)^2 \\ 
	&\quad \times \Big[  G^{(2)}_k(\varphi, {\bm q - \bm p}) -  G^{(2)}_k(\varphi, {\bm q})\Big] \,.
\end{align*}
The definition of the propagator $G^{(2)}_k$ can be found in \cref{eq:propagator-in-LPA}.
Since the propagators on the right-hand side of \cref{eq:LPA-flow-equation-for-kinetic} do not depend on $\Delta{\bar\Gamma}^{(2)}_k$ itself (as we work in \gls{LPA}), this flow equation is not a coupled differential equation and can therefore be integrated straightforwardly by inserting the solution for the effective potential~$U_k$ from \cref{eq:LPA-flow-equation} for a given set of parameters.
 
We emphasize that, for \gls{RG} scales above the lattice site decoupling scale $k^\star$ (i.e., in the \gls{UV} regime), the propagator $G^{(2)}_k(\varphi, {\bm q})$ becomes independent of the momenta and therefore the difference of the two propagators on the right-hand side of \cref{eq:LPA-flow-equation-for-kinetic} vanishes identically. 
In this regime, the kinetic term does not receive any quantum corrections in the \gls{RG} flow. 
This again reflects the exactness of \gls{LPA} at these scales, as discussed in \cref{sec:uv-regime}.

\Cref{eq:LPA-flow-equation-for-kinetic} can be used to estimate the uncertainty of \gls{LPA} in the intermediate regime where this approximation is not exact.
To be more specific, if we would find that the change of the momentum-dependent part of the two-point function relative to its classical form is exactly zero or at least very small for all momenta and field values, then \gls{LPA} can be expected to be a reasonable approximation for a determination of the effective potential. 
To quantify the uncertainty of \gls{LPA}, we therefore define 
\begin{align*}
	\numberthis\label{eq:relative-deviation}
	\mathcal{K}(\bm q) & = \max_{\substack{\varphi \geq \varphi_0(k=\epsilon_{\bm q})}}\left|\frac{\Delta\bar\Gamma^{(2)}_{k=\epsilon_{\bm q}}(\varphi, \bm q) - \epsilon^2_{\bm q}}{\epsilon^2_{\bm q}}\right| \,.
\end{align*}
This quantity represents the maximum relative deviation of the momentum-dependent part of the two-point function, as obtained in an \gls{LPA} flow for a given momentum~${\bm q}$, from the momentum dependence assumed in \gls{LPA}. 
The latter is nothing but the classical kinetic term.
Note that, in \cref{eq:relative-deviation}, we only take field values inside the physically relevant region into account, i.e., for $\varphi \geq \varphi_0(k)$. 
In the physically irrelevant region, i.e., for field values $\varphi < \varphi_0(k)$, the momentum-dependent part of the two-point function drastically changes since the potential becomes flat there. 
This would strongly dominate the relative deviation $\mathcal{K}$.
\begin{figure}
	\includegraphics{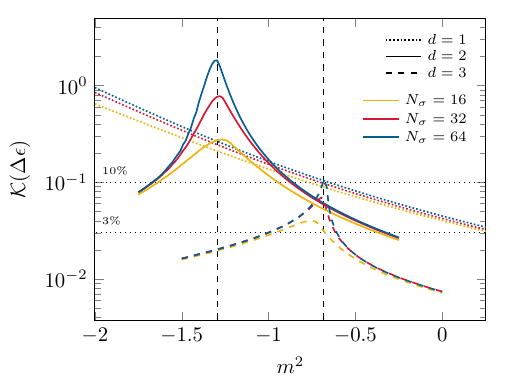}
	\caption{\label{fig:relative-deviation}The relative deviation of the fluctuation-induced kinetic term \eqref{eq:relative-deviation} from its classical form as a function of the bare mass $m^2$.}
\end{figure}

Because of the property \eqref{eq:ir-regulator-property} of the Litim regulator, only momentum-dependent contributions $\Delta\bar{\Gamma}^{(2)}_k(\varphi, \bm{q})$ with $\epsilon_{\bm{q}} < k$ are required to determine the next \gls{RG} step for the scale-dependent effective potential $U_k$, see \cref{eq:wetterich-equation-energy-specturm-decompositon-2}.
In other words, the evolution of the scale-dependent effective potential $U_k$ for $k\to 0$ is not directly affected by the parts of $\Delta\bar{\Gamma}^{(2)}_k(\varphi, \bm{q})$ with $\epsilon_{\bm{q}} > k$. 
This is also reflected in the decrease of the mode density $\Omega(k)$ as the \gls{IR} limit is approached, see \cref{fig:regimes}.
Therefore, to include only the regime of the \gls{RG} flow which affects the flow equation for the scale-dependent effective potential for a certain momentum $\bm q$ in \cref{eq:relative-deviation}, we evaluate $\Delta\bar{\Gamma}^{(2)}_k(\varphi, \bm{q})$ at $k = \epsilon_{\bm{q}}$.

It is important to emphasize that the quantity defined in \cref{eq:relative-deviation} serves solely as a measure to {\it estimate} the uncertainty of \gls{LPA}. 
It should be interpreted as follows: 
If the value of $\mathcal{K}(\bm q)$ is small, \gls{LPA} can be considered as a reliable approximation whereas no definitive statement can be made about the validity of \gls{LPA} for large~$\mathcal{K}(\bm q)$.

Since the \gls{RG} evolution of the two-point function depends on the solution for the effective potential~$U_k$ and, consequently, on the parameters that determine the bare action, namely $m^2$ and $\lambda$, we analyze the relative deviation~\eqref{eq:relative-deviation} as a function of $m^2$ while keeping $\lambda = 6$ fixed.
In \cref{fig:relative-deviation}, we show the relative deviation \eqref{eq:relative-deviation} evaluated on the mode associated with the lowest non-zero energy level  $\Delta\epsilon$ for one-, two- and three-dimensional systems with $N_{\sigma} = 16, 32, 64$ lattice sites in each direction.
Note that the modes associated with~$\Delta\epsilon$ are the only modes which remain ``active" in the \gls{RG} flow down to the \gls{IR} regime and can therefore significantly influence the evolution of the scale-dependent effective potential throughout the entire intermediate regime.
\begin{table}[b]
	\begin{ruledtabular}
		\setlength\extrarowheight{8pt}
		\begin{tabular}{l l l l l}
			$d$ & $N_\sigma$ & $\mathcal{K}(\Delta\epsilon) \geq 10\%$ & $\mathcal{K}(\Delta\epsilon) \geq 3\%$ & $m^2_{\mathrm{peak}}$ \\
			$2$ & $64$ & $m^2 \in [-1.659, -0.871]$ & $m^2 \in [-1.75, -0.325]$ & $-1.295$\\
			$3$ & $64$ & $m^2 \in \{-0.682\}$ & $m^2 \in [-1.0, -0.810]$ & $-0.682$\\
		\end{tabular}
	\end{ruledtabular}
	
	\caption{\label{tab:relative-deviation}%
		List of characteristic quantities of~$\mathcal{K}(\Delta\epsilon)$, i.e., the relative deviation of the fluctuation-induced kinetic term from its classical form, as extracted from \cref{fig:relative-deviation} for $N_{\sigma} = 64$.
	}
	
\end{table}
\begin{figure*}
	\subfloat[\label{fig:m2-scan-at-lowest-energy-level-phi0}%
	Magnetization.]{%
		\includegraphics{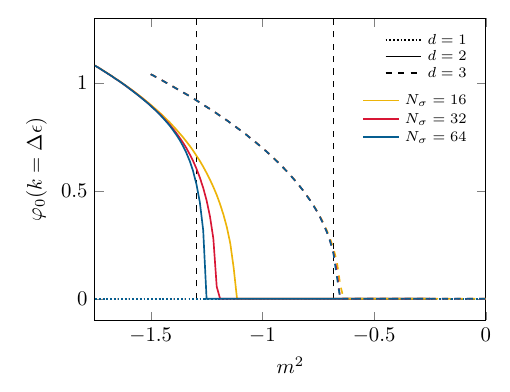}%
	}\hfill
	\subfloat[\label{fig:m2-scan-at-lowest-energy-level-correlation-length}%
	Correlation length.]{%
		\includegraphics{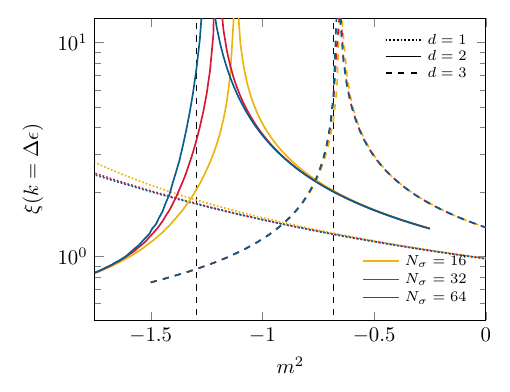}%
	}
	\caption{\label{fig:observables-at-lowest-energy-level}%
		Global minimum $\varphi_0(k)$ of the effective potential (left panel) and correlation length $\xi(k) = 1/m(k)$ (right panel) evaluated at the \gls{RG} scale $k = \Delta\epsilon$ for a fixed bare coupling~$\lambda=6$ as functions of the bare mass $m^2$ as obtained for different lattices sizes~$N_{\sigma}$ in $d=2$ and $d=3$ dimensions, respectively. Vertical lines mark the peak of the susceptibility on $N_\sigma=64$.
	}
\end{figure*}

We observe that the relative deviation is generally smaller for $d = 3$ than for $d = 2$ spacetime dimensions. 
This is in accordance with the general observation that the anomalous dimension at a critical fixed point increases in scalar field theories when the number of spacetime dimensions is decreased, see, e.g., Ref.~\cite{Pelissetto:2000ek} for a review. 
For example, the anomalous dimension at such a fixed point in $d=3$ is about one order of magnitude smaller than in $d=2$. 
Moreover, this observation with respect to the relative deviation is consistent with the fact that the critical exponents obtained in \gls{LPA} in $d=3$, where~$\eta=0$ by construction, already agree on the percent level with the world's best estimates, see, e.g., Refs.~\cite{Litim:2010tt,Pelissetto:2000ek,Litim:2002cf,Benitez:2009xg,Benitez:2011xx,Murgana:2023xrq}. 
At least close to a phase transition, a large anomalous dimension can therefore be considered an indication for the formation of a non-trivial momentum dependence of the two-point function. 
In any case, in both $d=2$ and~$d=3$, we observe peak-like structures that become sharper as $N_\sigma$ increases. 
For $N_\sigma=64$, the positions of these peaks are located at $m^2_{\mathrm{peak}} \approx  -1.295$ and $m^2_{\mathrm{peak}} \approx -0.682$ for $d=2$ and $d=3$, respectively. 
As we shall see below, these peak-like structures emerge close to the phase transition. 
For better guidance of the eye, we included horizontal lines in \cref{fig:relative-deviation} to indicate the regions in the $m^2$-plane where the relative deviation~$\mathcal{K}$ evaluated on the lowest non-trivial mode~$\Delta\epsilon$ is above $3\%$ and $10\%$, respectively. 
The precise values for the boundaries of these regions are listed in \cref{tab:relative-deviation} for $N_\sigma = 64$.

For completeness, we also show the scale-dependent global minimum $\varphi_0(k)$ as well as the scale-dependent correlation length $\xi(k) = 1/m(k)$ evaluated at the \gls{RG} scale $k = \Delta\epsilon$ as functions of $m^2$ in \cref{fig:observables-at-lowest-energy-level}. 
Note that, since we evaluated these quantities at a non-zero \gls{RG} scale, the effective potential $U_{k=\Delta\epsilon}$ need not be convex and the $Z(2)$ symmetry in the ground state is not yet necessarily restored at this scale. 
This explains the regions with a finite value of~$\varphi_0$ in \cref{fig:m2-scan-at-lowest-energy-level-phi0}.
The vertical dashed lines in both panels of \cref{fig:observables-at-lowest-energy-level} indicate the position of the peaks in the relative deviation~$\mathcal{K}$ for $N_{\sigma}=64$ in $d = 2$ and $d = 3$ spacetime dimensions, respectively, see \cref{fig:relative-deviation}.
Note that the peaks in the relative deviation do not coincide exactly with those of the correlation length, but approach each other as the spatial volume is increased.  This is a finite-size effect that will disappear in the thermodynamic limit,
where a non-analytic phase transition emerges, and indicates that the two-point function develops a non-trivial momentum dependence close to the phase transition.

In \cref{fig:relative-deviation}, we also show the relative deviation for $d=1$ spacetime dimensions. 
The corresponding partition function can be associated with the anharmonic oscillator in Quantum Mechanics. 
In this case, the curvature of the effective potential $U$ at its minimum is related to the energy difference between the two lowest levels of the system. 
Although the symmetry in the ground state is found to be restored in \gls{LPA}, as it should be, it has already been shown by comparison with exact results in Ref.~\cite{Kapoyannis:2000sp} that \gls{LPA} does not provide quantitative results for the energy difference of the two lowest-lying states for small values of the dimensionless coupling $\lambda/|m^2|^{3/2}$ with $m^2 < 0$ and $\lambda > 0$. 
This implies that \gls{LPA} does not allow to correctly recover the effective potential for classical potentials with a large potential barrier in $d=1$. 
This can be traced back to the relevance of instanton effects, which are not included in our current \gls{LPA} calculation~\cite{Kapoyannis:2000sp}. 
For sufficiently large values of the dimensionless coupling, \gls{LPA} yields results for the difference of the two lowest-lying energy levels which are in quantitative agreement with the exact results.
Note that this observation is in accordance with the behavior of the relative deviation \eqref{eq:relative-deviation} in \cref{fig:relative-deviation}. 
Indeed, we observe that the relative deviation $\mathcal{K}$ increases as $m^2$ is lowered for $d=1$.
This can be understood as follows: As we approach the limit of an infinitely negative value of $m^2$, the effective potential becomes arbitrarily flat and the correlation length increases accordingly, see \cref{fig:m2-scan-at-lowest-energy-level-correlation-length}. 
Note also that the non-trivial minimum $\varphi_0(\Delta \epsilon)$ of~$U$ must vanish for large enough volumes due to the Mermin-Wagner theorem.
As a consequence, {\it all} field values contribute to the relative deviation $\mathcal{K}$ as defined in \cref{eq:relative-deviation}, including those where the effective potential is very flat.

Finally, we would like to emphasize that all quantities shown in \cref{fig:relative-deviation,fig:observables-at-lowest-energy-level} carry an intrinsic dependence on the regulator $R_k$, irrespective of the fact that we did not solve the Wetterich equation exactly.
In fact, these quantities have been extracted from the \gls{RG} flow at a non-zero \gls{RG} scale $k$ which inherently depends on the choice of regulator. 
Note also that we choose $k = \Delta\epsilon$ since our regulator fulfills the property \eqref{eq:ir-regulator-property}. 
This renders the flow equation for the scale-dependent effective potential \eqref{eq:effective-potential-ir-regime} exact for $k < \Delta\epsilon$ which is only the case for a specific class of regulators.

In summary, we have defined an \gls{FRG}-intrinsic measure, the relative deviation of the momentum-dependent part of the two-point function from its classical counterpart, which allows us to estimate the uncertainty of \gls{LPA} in different regimes, see \cref{eq:relative-deviation}. 
Our analysis based on this measure indicates that \gls{LPA} tends to be more reliable the smaller the spacetime lattices, the higher the spacetime dimensions, and sufficiently far away from the critical regime. 
However, we stress that this criterion does not determine how a given relative deviation in the momentum-dependent part of the two-point function affects other physical observables. 
This question must be addressed by comparing our \gls{LPA} results with those obtained using the \gls{MC} approach.
\begin{figure}
	\includegraphics{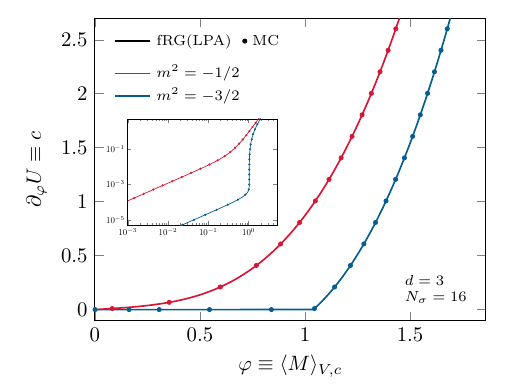}
	\caption{\label{fig:effective-potential}Field derivative of the effective potential~$U$ as a function of the field for \mbox{$m^2 = -1/2$} and $m^2=-3/2$ in three spacetime dimensions with $N_\sigma = 16$. 
	} 
\end{figure}
\subsection{Comparison of lattice \gls{MC} and lattice \gls{FRG}}\label{sec:comparison}
The present work aims at a quantitative comparison of two non-perturbative methods, lattice \gls{MC} and lattice \gls{FRG}, rather than at a study of phenomenological aspects of spin models.
For this comparison, we perform calculations over a wide range of the model parameters~$m$ and~$c$ while keeping the quartic coupling~$\lambda$ fixed. 

In addition to our \gls{FRG}-intrinsic analysis of the predictive power of \gls{LPA} in the previous subsection, a comparison of our lattice \gls{FRG} and lattice \gls{MC} results allows us to examine and quantify the limitations of \gls{LPA} in more detail.
\newcounter{subsubfigure}
\begin{figure*}
	\begingroup  
	\renewcommand{\thesubfigure}{\alph{subfigure}.\roman{subsubfigure}}
	
	\centering
	\setcounter{subsubfigure}{1}  
	\subfloat[\label{fig:phi4-comparison-LFRG-MC-1d-Ns-scan-large-mass-phi0}%
	Magnetization.]{%
	\includegraphics{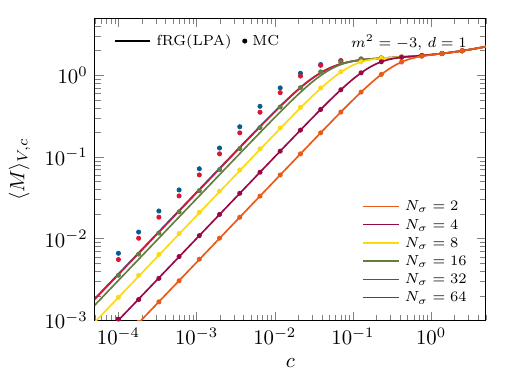}%
	}\hfill
	\setcounter{subfigure}{0}  
	\addtocounter{subsubfigure}{1}  
	\subfloat[\label{fig:phi4-comparison-LFRG-MC-1d-Ns-scan-large-mass-phi0}%
	Susceptibility.]{%
	\includegraphics{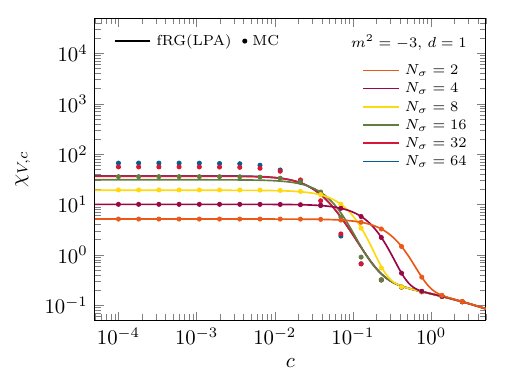}%
	}
	\setcounter{subfigure}{1}  
	\setcounter{subsubfigure}{1}  	
	\subfloat[\label{fig:phi4-comparison-LFRG-MC-1d-Ns-scan-small-mass-phi0}%
	Magnetization.]{%
	\includegraphics{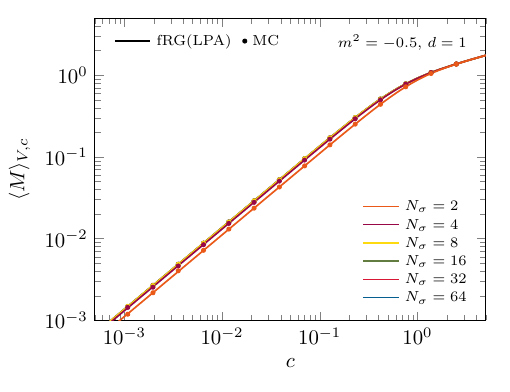}%
	}\hfill
	\setcounter{subfigure}{1}  
	\addtocounter{subsubfigure}{1}  
	\subfloat[\label{fig:phi4-comparison-LFRG-MC-1d-Ns-scan-small-mass-phi0}%
	Susceptibility.]{%
	\includegraphics{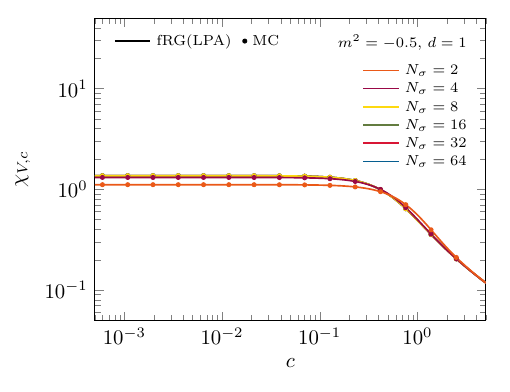}%
	}
	\caption{\label{fig:phi4-comparison-LFRG-MC-1d-Ns-scan}%
		Magnetization and susceptibility in one spacetime dimension 
		as a function of the external field~$c$ for two values 
		of the bare mass,  
		as obtained from lattice \gls{MC}~(dots) and lattice \gls{FRG}  (solid lines) calculations. 
		}
	\endgroup  
\end{figure*}
\subsubsection{Effective potential}
\label{subsubsec:effpotcomp}
In the \gls{FRG} approach we have direct access to the effective potential as it is the solution of the flow equation~\eqref{eq:LPA-flow-equation} in the \gls{IR} limit. 
For a lattice \gls{MC} computation of this potential, one may exploit the identity \eqref{eq:quantity-phi0}, i.e.,
\begin{align*}
	\partial_\varphi U(\langle M \rangle_{V, c}) = c \,.
\end{align*}
This equation relates the magnetization with the effective potential and can be used to obtain the latter by performing \gls{MC} calculations for different values of $c$. 
To be specific, by computing~$\langle M \rangle_{V, c}$ as a function of~$c$ and assuming that this relation can be inverted,
we find~$c=c(\langle M \rangle_{V, c})$. 
The effective potential in the absence of an external field can then be obtained as follows:
\begin{align*}
U(\varphi) = \int^{\varphi}_{\bar{\varphi}} {\rm d}\varphi^{\prime} c(\varphi^{\prime}) + \text{const.}\,,
\numberthis\label{eq:Ureconstruction}
\end{align*}
where we have used \eqref{eq:quantity-phi0} and the lower integration boundary is given by $\bar{\varphi}=\lim_{c\to 0}\langle M \rangle_{V, c}$. 

We emphasize that the effective potential is analytic and strictly convex in finite systems. 
In the thermodynamic limit, this is still the case in the absence of \gls{SSB}, where we have~$\bar{\varphi}=0$.
However, if the ground state is governed by \gls{SSB}, then the effective potential becomes non-analytic at~$\varphi = \bar{\varphi}=\lim_{c\to 0}\langle M \rangle_{V, c}$ and we have~$\partial_{\varphi}U(\varphi)=0$ for~$\varphi < \bar{\varphi}$, i.e., the effective potential is flat within this range of field values. 
These considerations imply that we can already analyze the shape of the effective potential by studying the dependence of the magnetization on~$c$. 
In particular, in our studies of finite systems, where the effective potential is analytic, the $c$-dependence of the magnetization and its susceptibility can be employed to detect regions in parameter space where the effective potential becomes flat over a finite range of field values, indicating \gls{SSB} in the thermodynamic limit. 
Recall that the susceptibility is determined by the inverse of the curvature of the effective potential, see \cref{eq:quantity-chi}. 
For example, a rapid increase in the susceptibility for small values of the external field~$c$ indicates the formation of a non-analyticity in the effective potential and the formation of a flat regime. 
We shall discuss this in more detail below. 

For illustration, we show the field derivative of the effective potential for $m^2 = -3/2$ and $m^2=-1/2$ in $d=3$ spacetime dimensions for $N_\sigma = 16$ as a function of the magnetization  in \cref{fig:effective-potential}.
These results imply that the associated effective potential is strictly convex in both cases and has a global minimum at $\varphi = 0$, reflecting the absence of \gls{SSB} in finite systems.
For $m^2 = -3/2$, however, our results for the field derivative of the effective potential indicate that the effective potential itself is already very flat in the region $0 \leq \varphi \lesssim 1$. 
From \cref{fig:effective-potential}, we can also deduce that the curvature of the effective potential undergoes a rapid change at the point where the potential becomes flat. 
This translates into a rapid change of the susceptibility~$\sim \partial\langle M \rangle_{V, c}/\partial c$ as a function of the external field~$c$, as we shall see below.
Following our discussion above, this behavior of the field derivative of the effective potential for~$m^2=-3/2$ can be considered a precursor for the formation of a phase with a finite magnetization in the thermodynamic limit. 
For $m^2=-1/2$, the situation is different. 
Indeed, we do not find that the potential develops a flat region in field space. 
Therefore, we expect the system to remain in the $Z(2)$-symmetric phase in the thermodynamic limit.  
In any case, for both values of $m^2$ , we find that the effective potential from our lattice \gls{FRG} study in \gls{LPA} agrees remarkably well with the results from our \gls{MC} calculations.

\subsubsection{Precursors of \gls{SSB} in finite systems}
\label{subsubsec:precursors}

Without explicitly considering the thermodynamic limit, we can already deduce from the behavior of the effective potential under a variation of~$N_\sigma$ (for a fixed lattice spacing) whether the ground state is governed by \gls{SSB} in the thermodynamic limit. 
As mentioned above, the behavior of the effective potential is also encoded in the magnetization as a function of the external field~$c$. 
To be specific, coming from large values of the external field~$c$, \gls{SSB} manifests itself as the formation of a plateau in the magnetization as~$c$ is decreased. 
For a system with a given set of model parameters in $d$ spacetime dimensions, we shall see that this plateau increases with increasing~$N_{\sigma}$ and eventually extends to~$c=0$, if the ground state is governed by \gls{SSB} in the thermodynamic limit.
The formation of such a plateau can therefore be regarded as a precursor of \gls{SSB} in finite systems. 
Of course, whether this plateau extends to~$c=0$ for~$N_{\sigma}\to \infty$ and thus truly indicates \gls{SSB} in the thermodynamic limit must always be analyzed by studying the scaling of the magnetization with~$N_{\sigma}$. 
We add that, in the presence of \gls{SSB} in the thermodynamic limit, the disappearance of the magnetization in a finite system for $c\to 0$ is a finite-volume effect. 

Let us now compare our results for the magentization and the susceptibility in different spacetime dimensions as obtained by our two non-perturbative approaches.
\begin{figure*}
	\begingroup  
	\renewcommand{\thesubfigure}{\alph{subfigure}.\roman{subsubfigure}}
	
	\centering
	\setcounter{subsubfigure}{1}  
	\subfloat[\label{fig:phi4-comparison-LFRG-MC-2d-Ns-scan-broken-phi0}%
	Magnetization.]{%
		\includegraphics{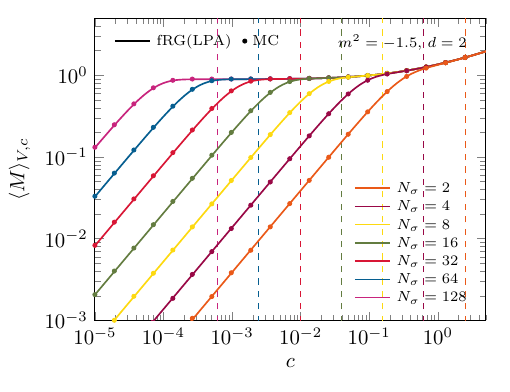}%
	}\hfill
	\setcounter{subfigure}{0}  
	\addtocounter{subsubfigure}{1}  
	\subfloat[\label{fig:phi4-comparison-LFRG-MC-2d-Ns-scan-broken-chi2}%
	Susceptibility.]{%
		\includegraphics{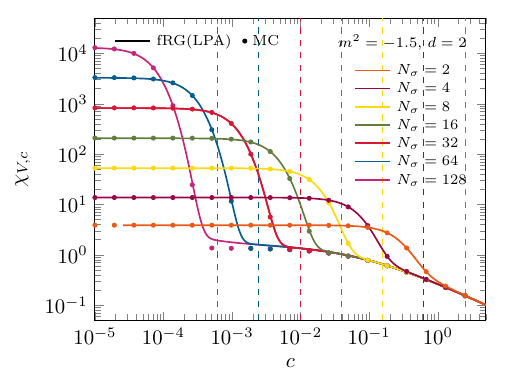}%
	}
	\setcounter{subfigure}{1}  
	\setcounter{subsubfigure}{1}  
	\subfloat[\label{fig:phi4-comparison-LFRG-MC-2d-Ns-scan-symmetric-phi0}%
	Magnetization.]{%
		\includegraphics{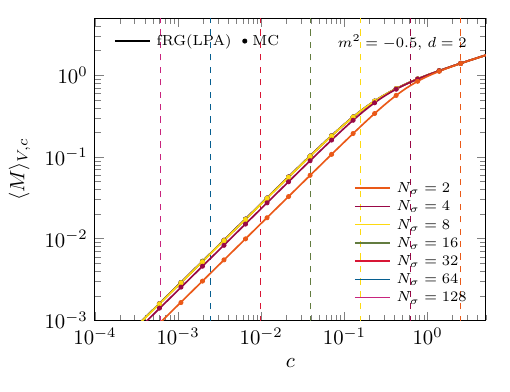}%
	}\hfill
	\setcounter{subfigure}{1}  
	\addtocounter{subsubfigure}{1}  
	\subfloat[\label{fig:phi4-comparison-LFRG-MC-2d-Ns-scan-symmetric-chi2}%
	Susceptibility.]{%
		\includegraphics{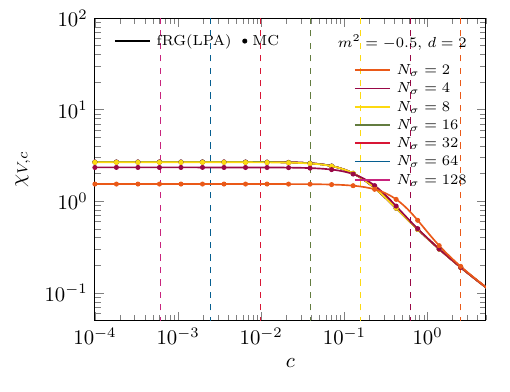}%
	}
	\caption{\label{fig:phi4-comparison-LFRG-MC-2d-Ns-scan}%
	Magnetization and susceptibility in two spacetime dimensions as a function of the external field~$c$ for two values of the bare mass, 
	as obtained from lattice \gls{MC}~(dots) and lattice \gls{FRG} (solid lines) calculations. 
	 The vertical dashed lines indicate the values of~$c$ at which the magnetization and susceptibility for the corresponding values of~$m^2$ and~$N_{\sigma}$ have been extracted for our analysis of the $m^2$-dependence of these quantities in \cref{fig:phi4-comparison-LFRG-MC-2d}, see main text for details. 
	}
	\endgroup  
\end{figure*}

\begin{figure*}
	\begingroup  
	\renewcommand{\thesubfigure}{\alph{subfigure}.\roman{subsubfigure}}
	
	\centering
	
	\setcounter{subsubfigure}{1}  
	\label{fig:test}
	\subfloat[\label{fig:phi4-comparison-LFRG-MC-3d-Ns-scan-broken-phi0}%
	Magnetization.]{%
		\includegraphics{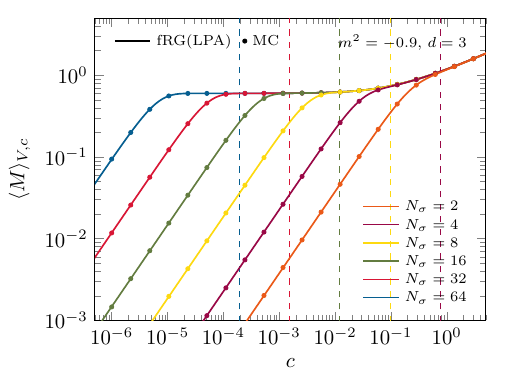}%
	}\hfill
	\setcounter{subfigure}{0}  
	\addtocounter{subsubfigure}{1}  
	\subfloat[\label{fig:phi4-comparison-LFRG-MC-3d-Ns-scan-broken-chi2}%
	Susceptibility.]{%
		\includegraphics{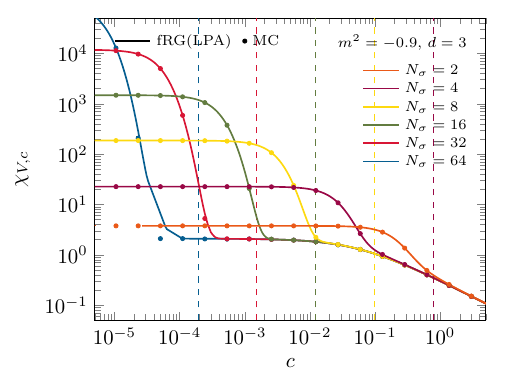}%
	}
	\setcounter{subfigure}{1}  
	\setcounter{subsubfigure}{1}  
	\subfloat[\label{fig:phi4-comparison-LFRG-MC-3d-Ns-scan-symmetric-phi0}%
	Magnetization.]{%
		\includegraphics{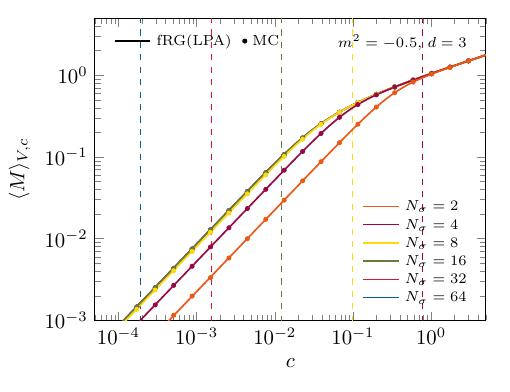}%
	}\hfill
	\setcounter{subfigure}{1}  
	\addtocounter{subsubfigure}{1}  
	\subfloat[\label{fig:phi4-comparison-LFRG-MC-3d-Ns-scan-symmetric-chi2}%
	Susceptibility.]{%
		\includegraphics{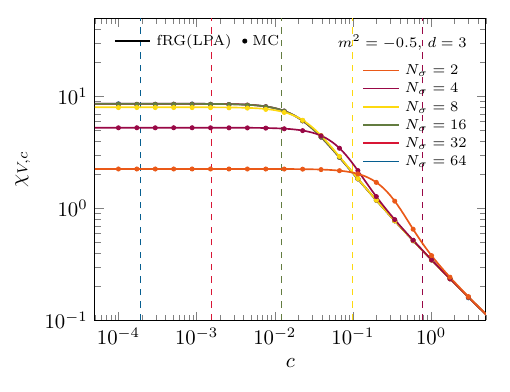}%
	}
	\caption{\label{fig:phi4-comparison-LFRG-MC-3d-Ns-scan}%
Magnetization and susceptibility in three spacetime dimensions as a function of the external field~$c$ for two values of the bare mass, as obtained from lattice \gls{MC}~(dots) and lattice \gls{FRG} (solid lines) calculations. 
The vertical dashed lines indicate the values of~$c$ at which the magnetization and susceptibility for the corresponding values of~$m^2$ and~$N_{\sigma}$ have been extracted for our analysis of the $m^2$-dependence of these quantities in \cref{fig:phi4-comparison-LFRG-MC-3d}, see main text for details.
}	
	\endgroup  
\end{figure*}
\paragraph{One dimension.}
For~$m^2>0$, this system corresponds to the anharmonic oscillator in Quantum Mechanics.
The case with~$m^2<0$, which we consider from here on, is a model to study tunneling in Quantum Mechanics. 
In any case, the relation to quantum-mechanical one-particle systems already indicates that \gls{SSB} cannot occur in $d=1$, which we also find in our present study. 
Note that this is correct regardless of our choice of model parameter values. 
Accordingly, the magnetization in one dimension must vanish when we consider the limit~$c\to 0$ after the limit $N_{\sigma}\to\infty$ has been taken. 

In \cref{fig:phi4-comparison-LFRG-MC-1d-Ns-scan}, we show the magnetization and the susceptibility as a function of~$c$ for various values of~$N_{\sigma}$ and two values for~$m^2<0$. 
As explained above, the quartic coupling has been set to the same value $\lambda =6$ in the two cases. 
For all considered values of~$N_{\sigma}$ and~$m^2$, we do not observe the formation of a plateau in the magnetization as a function of the external field~$c$. 
In fact, the magnetization tends to zero as we decrease~$c$ while the susceptibility remains finite. 
Moreover, in accordance with our \gls{FRG}-intrinsic analysis of the reliability of \gls{LPA} in \cref{sec:assessing-lpa}, we find that the deviation of our lattice \gls{FRG} results from the lattice \gls{MC} results increases with decreasing~$m^2$.
This can be traced back to the fact that an accurate resolution of the momentum dependence of the correlation functions becomes very relevant for a quantitatively correct description of tunneling through a (high) potential barrier. 
In any case, while inaccurate on a quantitative level for increasing~$N_{\sigma}$, the disappearance of the magnetization for~$c\to 0$ in the thermodynamic limit is still observed in \gls{LPA}.

We close the discussion of the one-dimensional case by adding that the lattice \gls{MC} and lattice \gls{FRG} results are overall in excellent agreement for small values of~$N_{\sigma}$. 
This is true regardless of the dimension of the system, see also below.
Of course, this does not come unexpected as \gls{LPA} becomes exact for~$N_{\sigma}=1$ which corresponds to the case of a zero-dimensional quantum field theory~\cite{Koenigstein:2021syz,Koenigstein:2021rxj,Steil:2021cbu,Zorbach:2024rre}.

\paragraph{Two dimensions.}  
In \cref{fig:phi4-comparison-LFRG-MC-2d-Ns-scan}, the magnetization and the susceptibility are shown as functions of the external field~$c$ for~$m^2=-1.5$ and~$m^2=-0.5$ as obtained for various values of~$N_{\sigma}$ in two spacetime dimensions. 
The quartic coupling is the same in both cases. 
We readily observe that the lattice \gls{MC} and lattice \gls{FRG} results for the magnetization are in good agreement. 
The two values selected for the parameter~$m^2$ are associated with qualitatively different situations in the thermodynamic limit, as we shall see next. 

For~$m^2=-1.5$, we observe the formation of a plateau in the magnetization as a function of the external field~$c$, which increases continuously as we increase~$N_{\sigma}$. 
Thus, for this value of~$m^2$ we expect the system to be in the symmetry broken phase where the ground state is governed by spontaneous $Z(2)$ symmetry breaking in the thermodynamic limit for~\mbox{$c\to 0$}. 
Our results make apparent that the order of the limits~$c\to 0$ and~$N_{\sigma}\to \infty$ do not commute.
In fact, to obtain a finite magnetization in the thermodynamic limit, we have to take the limit $c \to 0$ {\it after} the thermodynamic limit, $N_{\sigma} \to \infty$, see also \cref{eq:magnetization}.

For~$m^2=-0.5$, we do not observe the formation of a plateau in the magnetization, even for large values of~$N_{\sigma}$. 
For increasing~$N_{\sigma}$, we rather find that the results from both methods converge to a continuous function which tends to zero for \mbox{$c\to 0$}.
Consequently, we expect the system to be in the symmetric phase in the thermodynamic limit for this value of~$m^2$. 

Let us now consider the susceptibility, which is a more sensitive probe for the detection of differences between our lattice \gls{FRG} and the lattice \gls{MC} results, since it corresponds to the derivative of the magnetization with respect to the external field~$c$ and measures fluctuations. 
Still, we observe that the results for the susceptibility agree well for~$m^2=-0.5$. 
However, deviations are found for~$m^2=-1.5$ as~$N_{\sigma}$ increases.  
Note that, for this value of~$m^2$, the system is in the symmetry broken phase but still not far away from the phase transition in~$m^2$, see our discussion of phase transitions below. 
Since our lattice \gls{FRG} calculations are based on \gls{LPA}, these deviations of the lattice \gls{FRG} results from the lattice \gls{MC} results already hint at the importance of a non-trivial momentum dependence in the correlation functions, which become increasingly relevant close to the phase transition. 
We shall come back to this aspect below, as we would first like to discuss characteristic features of the susceptibility in finite systems.

In our results for the susceptibility in the symmetry broken phase approaching the thermodynamic limit, we observe the formation of two plateaus, one appearing for very small values of~$c$ and the other for small but not too small values of $c$, see \cref{fig:phi4-comparison-LFRG-MC-2d-Ns-scan-broken-chi2} for an illustration. 
The latter plateau determines the value of the susceptibility in the thermodynamic limit for~$c\to 0$. 
In fact, this plateau extends to smaller values of~$c$ as~$N_{\sigma}$ increases, and would end in a finite value if we take the limit~$c\to 0$ after the thermodynamic limit. 

To understand the second plateau in the susceptibility, which appears at small values of~$c$ in {\it finite} systems, it is instructive to recall how the effective potential can be reconstructed from the dependence of the external field on the magnetization, $c=c(\langle M \rangle_{V, c})$. 
Note that~$c$ increases strictly monotonically with $\langle M \rangle_{V, c}$ and we have~$c\to 0$ for~$\langle M \rangle_{V, c}\to 0$ in finite systems. 
Following our discussion of Eq.~\eqref{eq:Ureconstruction}, the function $c(\langle M \rangle_{V, c})$ can be identified with the field derivative of the effective potential,~$\partial_{\varphi}U$. 
Accordingly, the susceptibility~$\sim \partial\langle M \rangle_{V, c}/\partial c$ can be related to the inverse of the curvature of the effective potential, $1/(\partial_{\varphi}^2 U)$.
Starting from large values of~$c$, the rapid increase in the susceptibility to large values, accompanied by the formation of a plateau at small values of~$c$, corresponds to a flattening of the effective potential for field values smaller than the one associated with the non-trivial minimum of the effective potential in the thermodynamic limit. 
Since convexity requires that the curvature of the effective potential must be zero for~$|\varphi| <  \langle M \rangle$ in the symmetry broken phase in the thermodynamic limit, the plateau of the susceptibility at small values of~$c$ must increase as~$N_{\sigma}$ increases.
This is exactly what we observe in \cref{fig:phi4-comparison-LFRG-MC-2d-Ns-scan-broken-chi2}. 

For values of~$m^2$ associated with a magnetization that vanishes in the thermodynamic limit for~$c\to 0$, the curvature of the corresponding effective potential is finite and positive for all field values. 
As $c$ is decreased, we therefore observe that the susceptibility only develops a single plateau in this case, see \cref{fig:phi4-comparison-LFRG-MC-2d-Ns-scan-symmetric-chi2}. 
The height of this plateau determines the value of the susceptibility in the thermodynamic limit.

\paragraph{Three dimensions.}  
Now we turn to the three-dimensional case which is most relevant from the standpoint of an analysis of finite-temperature phase transitions in $3+1$-dimensional spacetime. 

In \cref{fig:phi4-comparison-LFRG-MC-3d-Ns-scan}, the magnetization and the susceptibility are shown as functions of the external field $c$ for various values of~$N_{\sigma}$. 
As for the two-dimensional system, we show results for two values of the parameter~$m^2$, one of which, $m^2=-0.9$, is associated with the symmetry broken phase in the thermodynamic limit, while the other, $m^2=-0.5$, is associated with the symmetric phase in the thermodynamic limit. 
Qualitatively, the magnetizations and susceptibilities associated with the two phases behave as their analogues in two dimensions.
In fact, for $m^2=-0.9$, we find the formation of a plateau in the magnetization as a function of the external field~$c$. 
As we increase~$N_{\sigma}$ this plateau grows continuously and is expected to extend to $c=0$ for~$N_{\sigma}\to \infty$.
This behavior signals that the magnetization remains finite in the thermodynamic limit, even in the absence of an external field. 
The susceptibility exhibits two plateaus as also observed for the two-dimensional system: one determining the susceptibility in the thermodynamic limit for~$c\to 0$, and one indicating that the effective potential in the thermodynamic limit becomes flat for field values smaller than the one of the minimum.
Again, we also observe that the limits $c\to 0$ and~$N_{\sigma}\to \infty$ do not commute in the symmetry broken phase.

Let us now come to the case with $m^2 = -0.5$. 
Here, we do not observe the formation of a plateau in the magnetization as a function of the external field~$c$ as we increase~$N_{\sigma}$. 
In fact, as we increase~$N_{\sigma}$, we find that the magnetization converges to a continuous function which tends to zero for vanishing~$c$. 
The susceptibility exhibits a similar convergent behavior and, as for the two-dimensional system, develops only a single plateau and approaches a finite value for~$c\to 0$.  
This behavior of the magnetization and the susceptibility indicates that the $Z(2)$ symmetry is restored for $m^2 = -0.5$ in the thermodynamic limit.

Overall, we find that the lattice \gls{FRG} and lattice \gls{MC} results are in remarkable agreement, given the fact that the lattice \gls{FRG} calculations are based on \gls{LPA}. 
The reader may note apparent deviations of the lattice \gls{FRG} results from the lattice \gls{MC} results in the susceptibility for large values of~$N_{\sigma}$ and those values of~$c$ associated with the regime between the two plateaus.  
We emphasize that these deviations are only numerical artefacts of the lattice \gls{FRG} calculations, which can in principle be removed by increasing the resolution of the grid in field space. 
For details on the numerical setup used for the lattice \gls{FRG} calculations we refer the reader to Appendix~\ref{sec:solving-the-flow-equation}. 

Finally, looking at our results for the magnetization and susceptibility in different numbers of spacetime dimensions, we find that the results obtained from lattice \gls{FRG} in \gls{LPA} and lattice \gls{MC} are not only consistent on a qualitative level, but also become successively more consistent on a quantitative level as the number of dimensions increases. 
Without presenting numerical results here, we add that this is indeed confirmed by calculations of the magnetization and susceptibility in four spacetime dimensions.
\begin{figure*}
	\subfloat[\label{fig:phi4-comparison-LFRG-MC-2d-phi0}%
	Magnetization.]{%
		\includegraphics{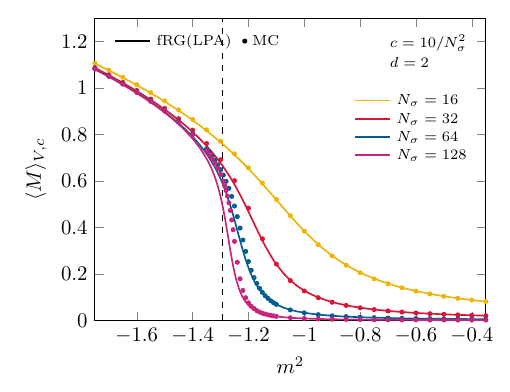}%
	}\hfill
	\subfloat[\label{fig:phi4-comparison-LFRG-MC-2d-chi2}%
	Susceptibility.]{%
		\includegraphics{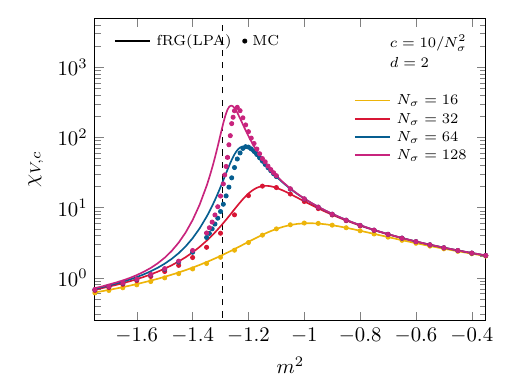}%
	}
	\caption{\label{fig:phi4-comparison-LFRG-MC-2d}%
		Magnetization and chiral susceptibility in two spacetime dimensions as a function of the bare mass for a fixed explicit symmetry breaking~$c$, as obtained from lattice calculations~(dots) and lattice-FRG calculations (solid lines).
		The vertical lines in the two panels indicate the position where our  \gls{FRG}-intrinsic analysis of the predictive power of \gls{LPA} suggests the largest deviations from the exact solution, see \cref{sec:assessing-lpa}. 
	}
\end{figure*}
\subsubsection{Phase transitions}
Above, we have discussed precursors of \gls{SSB} in finite systems. 
In the following, we shall study the approach to phase transitions in two and three dimensions in the thermodynamic limit. 
This requires a calculation of the magnetization and the susceptibility as a function of the parameter~$m^2$, which mimics the temperature in a thermodynamic system in one dimension higher.

However, before actually analyzing the transition from the symmetry broken to the symmetric phase, it is necessary to discuss briefly finite-volume effects which are present for small values of the external field~$c$. 
Such effects become most pronounced for values of~$m^2$ close to the phase transition (or crossover for finite~$c$), as the correlation length becomes large in this regime. 
For example, this is the case for $m^2=-1.5$ in two dimensions, see \cref{fig:phi4-comparison-LFRG-MC-2d-Ns-scan-broken-phi0,fig:phi4-comparison-LFRG-MC-2d-Ns-scan-broken-chi2}, and for $m^2=-0.9$ in three dimensions, see \cref{fig:phi4-comparison-LFRG-MC-3d-Ns-scan-broken-phi0,fig:phi4-comparison-LFRG-MC-3d-Ns-scan-broken-chi2}.
\begin{figure*}
	\subfloat[\label{fig:phi4-comparison-LFRG-MC-3d-phi0}%
	Magnetization.]{%
		\includegraphics{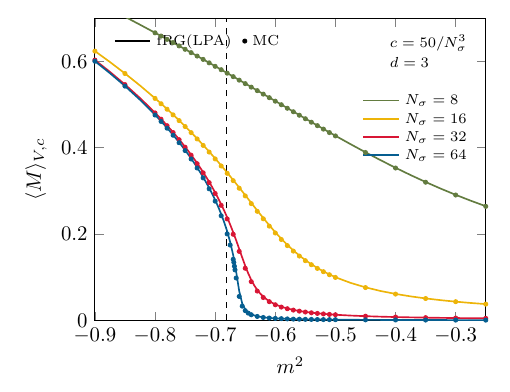}%
	}\hfill
	\subfloat[\label{fig:phi4-comparison-LFRG-MC-3d-chi2}%
	Susceptibility.]{%
		\includegraphics{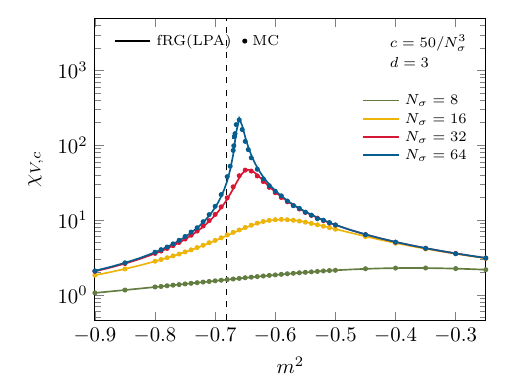}%
	}
	\caption{\label{fig:phi4-comparison-LFRG-MC-3d}%
		Magnetization and chiral susceptibility in three spacetime dimensions as a function of the bare mass for a fixed explicit symmetry breaking~$c$, as obtained from lattice calculations~(dots) and lattice-FRG calculations (solid lines). 
		The vertical lines in the two panels indicate the position where our  \gls{FRG}-intrinsic analysis of the predictive power of \gls{LPA} suggests the largest deviations from the exact solution, see \cref{sec:assessing-lpa}. 
	}
\end{figure*}

To mimick properties of the system in the thermodynamic limit, we need to suppress finite-volume effects. 
This can be done by determining a specific value of the external field, $c_{\star}(N_{\sigma})$, such that $\langle M \rangle_{V, c} \approx \langle M \rangle_{c}$ and $\chi_{V, c} \approx \chi_{c}$ for all $c \geq c_{\star}(N_{\sigma})$ for a given~$d$. 
Here, $\langle M \rangle_{c}$ and $\chi_{c}$ are the values of the magnetization and susceptibility in the thermodynamic limit in the presence of the external field~$c$.
Moreover, $c_{\star}(N_{\sigma})$ should be chosen such that it is as small as possible and vanishes in the limit $N_{\sigma}\to \infty$.
With this quantity at hand, we have
\begin{align*}
	\numberthis\label{eq:magnetization-2}
	\langle M \rangle = \lim_{N_{\sigma}\to \infty} \langle M \rangle_{V, c_{\star}(N_{\sigma})}
\end{align*}
for a given~$d$. 
Note that we shall determine~$c_{\star}$ such that it does not depend on the bare parameters~$m^2$ and~$\lambda$. 

To find an estimate for~$c_{\star}$, we consider the susceptibility and determine the value of~$c$ at which finite-volume effects set in. 
For~$d=2$, we find that $c_{\star}(N_{\sigma}) = 10 / (aN_{\sigma})^2$ is an appropriate choice. 
For illustration, we show the corresponding values of~$c_{\star}$ as vertical dashed lines in \cref{fig:phi4-comparison-LFRG-MC-2d-Ns-scan-broken-phi0,fig:phi4-comparison-LFRG-MC-2d-Ns-scan-broken-chi2}. 
For~$d=3$, we obtain $c_{\star}(N_{\sigma}) = 50 / (aN_{\sigma})^3$, see the vertical dashed lines in \cref{fig:phi4-comparison-LFRG-MC-3d-Ns-scan-broken-phi0,fig:phi4-comparison-LFRG-MC-3d-Ns-scan-broken-chi2}.

With~$c_{\star}(N_{\sigma})$ for two and three spacetime dimensions at hand, we can compute the magnetization and susceptibility as a function of the squared bare mass parameter~$m^2$ to detect the formation of phase transitions in the thermodynamic limit.
However, strictly speaking, phase transitions cannot occur in a finite system. 
The search for the emergence of non-analyticities associated with phase transitions therefore requires an analysis of the scaling behavior of the magnetization and susceptibility with~$N_{\sigma}$. 

A detailed scaling analysis is beyond the scope of the present work. 
We shall only illustrate the scaling behavior of the magnetization and susceptibility in \cref{fig:phi4-comparison-LFRG-MC-2d,fig:phi4-comparison-LFRG-MC-3d} for~$d=2$ and $d=3$, respectively. 
In these figures, the magnetization $\langle M \rangle_{V, c}$ and the susceptibility $\chi_{V, c}$ are shown as functions of the squared (bare) mass $m^2$ for various values of~$N_{\sigma}$. 
As explained above, the values of the external field~$c$ have been chosen such that $c=c_{\star}(N_{\sigma}) = 10 / (a N_{\sigma})^2$ for~$d=2$ and $c=c_{\star}(N_{\sigma})=50/(a N_{\sigma})^3$ for~$d=3$. 
The vertical lines  in \cref{fig:phi4-comparison-LFRG-MC-2d,fig:phi4-comparison-LFRG-MC-3d} represent the values $m^2_{\mathrm{peak}} \approx -1.295$ for $d=2$ and $m^2_{\mathrm{peak}} \approx -0.682$ for $d=3$, respectively.
These values correspond to the values of the bare mass where the \gls{FRG}-intrinsic analysis of the predictive power of \gls{LPA} suggests the largest deviations from the exact solution, see \cref{sec:assessing-lpa}. 
Note that these values should not be confused with the critical bare mass value associated with the phase transition in the thermodynamic limit.

For~$d=2$ and~$d=3$, we observe a behavior of the magnetization and susceptibility in \cref{fig:phi4-comparison-LFRG-MC-2d,fig:phi4-comparison-LFRG-MC-3d}, which is indicative of a second-order phase transition: the magnetization develops a pronounced kink as~$N_{\sigma}$ increases, and the susceptibility increases with~$N_{\sigma}$, indicating the formation of a divergence. 

Comparing the results for the magnetization and susceptibility from our lattice \gls{FRG} studies in \gls{LPA} with those from our lattice \gls{MC} calculations, we find excellent agreement for small~$N_{\sigma}$. 
In the symmetric phase, this appears to hold even for larger values of~$N_{\sigma}$. 
However, for~$d=2$, significant deviations appear in the symmetry broken phase, see~\cref{app:d2deviations} for a more detailed analysis.
This observation is in accordance with our \gls{FRG}-intrinsic analysis of the predictive power of \gls{LPA} in \cref{sec:assessing-lpa}.
In fact, this analysis already indicates that the deviations of the lattice \gls{FRG} results in \gls{LPA} from the exact solution should be expected to be larger in~$d=2$ than in~$d=3$.
Note that the deviations in the magnetization and susceptibility are indeed maximal around $m^2=m^2_{\mathrm{peak}}$, as predicted by our \gls{FRG}-intrinsic analysis. 
Apparently, $m^2_{\mathrm{peak}}$ is close to the phase transition in both $d=2$ and~$d=3$.

We conclude this section by adding that the good agreement between the results of our lattice \gls{FRG} calculations in \gls{LPA} and lattice \gls{MC} studies in $d=3$ is also not unexpected from a more general standpoint.
In fact, the anomalous dimension~$\eta$, which can be viewed as a measure of the relevance of non-trivial momentum dependences in correlation functions, is small at the phase transition in $d=3$,~$\eta\approx 0.036$, see, e.g.,  Refs.~\cite{Pelissetto:2000ek,Benitez:2009xg,Benitez:2011xx}. 
In \gls{LPA}, we have~$\eta=0$ by construction, regardless of the dimension of the system.  
It is then also reasonable that the situation is different in two spacetime dimensions. 
There, the anomalous dimension is about an order of magnitude larger than in three spacetime dimensions~\cite{Pelissetto:2000ek}, indicating the relevance of non-trivial momentum structures in, e.g., the propagator.
From a more phenomenological standpoint, the potential relevance of non-trivial momentum dependences close to the phase transition appears reasonable since the particles associated with our quantum field become massless at the phase transition. 
Away from the transition, both in the symmetry broken and symmetric phase, the masses of these particles are finite which suppresses momentum dependences in correlation functions.
This is indeed confirmed by the particularly good agreement between our results of lattice \gls{MC} and lattice \gls{FRG} in \gls{LPA} away from the phase transition.

\section{Conclusion}\label{sec:conclusion}
In the present work, we have introduced a framework for a direct comparison of lattice \gls{MC} and lattice \gls{FRG} studies on
finite volumes and at fixed lattice spacing,  
thus avoiding any non-trivial parameter matching between the two. 
In particular, this allows for a clear analysis of a wide range of artefacts, such as cutoff, finite-volume and truncation effects. 

As a first application of our framework, we have considered a scalar $Z(2)$ theory in various spacetime dimensions and provided detailed comparisons for the magnetization, the susceptibility, and phase transitions. 
For a given size of the spacetime lattice, and at fixed lattice spacing, the lattice \gls{MC} results contain only statistical errors, which for these simple systems can be made arbitrarily small. 
In such a situation, our framework is ideally suited to analyze the predictive power of trunctaions entering the computations within the \gls{FRG} approach. 
In the present work, we have demonstrated this by comparing lattice \gls{MC} results with results from lattice \gls{FRG} calculations in \gls{LPA}. 
Within the \gls{FRG} approach, this is the simplest approximation that already takes into account fluctuation effects. 
Indeed, this approximation of the effective action at leading order in a derivative expansion has been widely used in the past and is still frequently used in various research fields.

For a small number of lattice sites, we have found that the lattice \gls{FRG} results in \gls{LPA} are in excellent agreement with our \gls{MC} results, regardless of the number of spacetime dimensions. 
We have shown that this follows from the fact that \gls{LPA} becomes exact in the limit of a lattice consisting of only a single spacetime point. 
By increasing the number of lattice sites, we have observed that the results for the magnetization and susceptibility from the two methods start to deviate in regimes associated with a small mass of the field, e.g., close to the phase transition in two and three spacetime dimensions, but still remain in agreement at a qualitative level. 
The size of the aforementioned deviations depends on the number of spacetime dimensions. 
In general, however, our results indicate that the deviations become smaller as the number of spacetime dimensions increases, such that the lattice \gls{FRG} results in \gls{LPA} and the lattice \gls{MC} results become successively more consistent on a quantitative level.
In fact, while the deviations in the magnetization and especially in the susceptibility are still significant around the phase transition in two spacetime dimensions, the lattice \gls{FRG} and lattice \gls{MC} results show remarkable agreement in three spacetime dimensions, away from the phase transition but also close to it. 
Given the simplicity of \gls{LPA}, this is indeed impressive. 
Our analysis indicates that this can be traced back to the fact that non-trivial momentum dependences in the correlation functions become less relevant in higher dimensions, at least with respect to calculations of the magnetization and susceptibility.
This observation is consistent with the anomalous dimension at the phase transition being one order of magnitude smaller in three spacetime dimensions than in two spacetime dimensions. 

In addition to testing the predictive power of \gls{FRG} approximation schemes, as exemplified in our present work, it may be beneficial for lattice \gls{MC} studies to exploit the fact that lattice \gls{FRG} calculations can be used to track the scaling behavior of observables from very small lattices up to the thermodynamic limit, as well as the approach to the continuum limit. 
For example, provided that the results of both methods are found to agree well over a range of lattice sizes, our lattice \gls{FRG} approach can be used to guide extrapolations of lattice \gls{MC} data. 
This may be relevant for theories with fermions or for tests of methods developed to surmount the sign problem at finite density. 
Conversely, the very good agreement of our lattice \gls{FRG} results in \gls{LPA} and lattice \gls{MC} results over a wide range of lattice sizes indicates that large lattices may be required to resolve the effect of non-trivial momentum dependences of correlation functions on observables, e.g., in the critical regime.

In general, the opportunity to make clear and meaningful comparisons of lattice \gls{MC} and \gls{FRG} studies offers great potential, as it may lead to cross-fertilization and improvements on both sides in the future. 
\begin{figure*}
	\subfloat[\label{fig:phi4-comparison-LFRG-MC-2d-Ns-scan-broken-critical-regime-phi0}%
	Magnetization.]{%
		\includegraphics{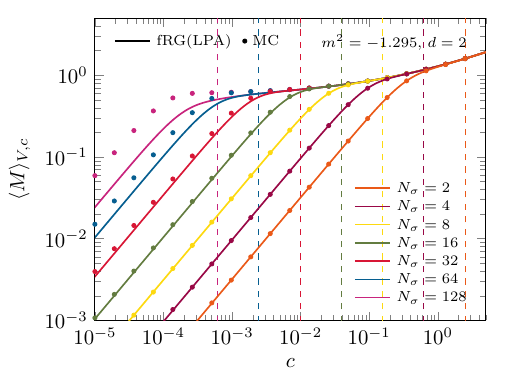}%
	}\hfill
	\subfloat[\label{fig:phi4-comparison-LFRG-MC-2d-Ns-scan-broken-critical-regime-chi2}%
	Susceptibility.]{%
		\includegraphics{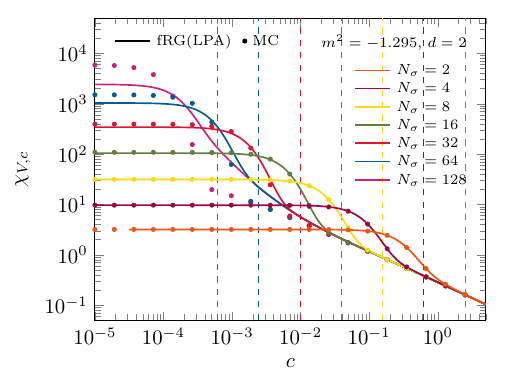}%
	}
	\caption{\label{fig:phi4-comparison-LFRG-MC-2d-Ns-scan-critical-regime}%
			Magnetization and susceptibility in two spacetime dimensions from lattice \gls{MC}~(dots) and lattice \gls{FRG} (solid lines) calculations as a function of the external field~$c$ at $m^2=m^2_{\mathrm{peak}} \approx -1.295$. At this value of~$m^2$, the greatest discrepancy of the lattice \gls{FRG} results in \gls{LPA} and lattice \gls{MC} results is observed, see vertical dashed line in  \cref{fig:phi4-comparison-LFRG-MC-2d}.
			The vertical dashed lines again indicate the values of~$c$ at which the magnetization and susceptibility for the corresponding values of~$m^2$ and~$N_{\sigma}$ have been extracted for our analysis of the $m^2$-dependence of these quantities in \cref{fig:phi4-comparison-LFRG-MC-2d}.
			}
\end{figure*}

\acknowledgements
We thank A.~Koenigstein, J.~M. Pawlowski, and J.~Stoll for discussions and comments on the manuscript. 
This work is supported by the \textit{Deutsche Forschungsgemeinschaft} (DFG, German Research Foundation) through the CRC-TR 211 ``Strong-interaction matter under extreme conditions'' -- project number 315477589 -- TRR 211 and by the State of Hesse within the Research Cluster ELEMENTS (Project No. 500/10.006). 
\appendix
\section{Numerical implementation of the \gls{FRG} flow equation of the effective potential}\label{sec:solving-the-flow-equation}
To solve the flow equation for the effective potential, which is a highly non-linear diffusion equation, we have brought it into a conservative form by taking a field derivative of it~\cite{Grossi:2019urj,Koenigstein:2021syz}. 
The resulting equation can then be solved by using a so-called finite-volume method based on the \gls{KT} scheme, see Ref.~\cite{KTO2-0}. 
In the present work, we have used the same semi-discrete implementation as described in Refs.~\cite{Koenigstein:2021syz,Zorbach:2024rre}. 
For the numerical time stepper, we have used \texttt{solve\_ivp} with \texttt{LSODA} and $a_{\mathrm{tol}} = r_{\mathrm{tol}} = 10^{-14}$ for its absolute and relative tolerances, respectively, if not stated otherwise. 
To obtain the numerical results shown in \cref{fig:regimes,fig:relative-deviation,fig:observables-at-lowest-energy-level,fig:effective-potential}, we have moreover used an equidistant grid in field space with spacing $\Delta \varphi = 0.001$, while we have used $\Delta \varphi = 0.0001$ to obtain the results shown in all other figures. 
For the maximal field value, we have used $\varphi_{\mathrm{max}} = 3$ for $d=2,3$ and $\varphi_{\mathrm{max}} = 5$ for $d=1$. 
At the boundaries in field space, we have followed Ref.~\cite{Koenigstein:2021syz} and employed a linear extrapolation at $\varphi=0$ and $\varphi = \varphi_{\mathrm{max}}$. 
For the initial \gls{RG} scale~$\Lambda$, we have used~$\Lambda=100/a$ in all numerical calculations, which effectively removes the dependence of our results from this scale. 
This is in accordance with our discussion in \cref{sec:lfrg} where we show that the limit~$\Lambda\to\infty$ can be taken for a given finite lattice spacing~$a$.
In the \gls{IR} regime, we have always stopped the \gls{RG} flow at~$k/\Lambda=k_{\text{IR}}/\Lambda=10^{-12}$. 

\section{External field dependence close to the phase transition in two spacetime dimensions}
\label{app:d2deviations}
In \cref{subsubsec:precursors}, we have discussed the dependence of the magnetization and susceptibility on the external field~$c$ for two values of the squared (bare) mass parameter~$m^2$ in two spacetime dimensions, see \cref{fig:phi4-comparison-LFRG-MC-2d-Ns-scan}. 
The results in this figure show that the lattice \gls{MC} results and the lattice \gls{FRG} results in \gls{LPA} agree well for both values of~$m^2$. 
Deviations in the susceptibility emerge only for very large lattices close to the phase transition. 
However, this observation is somewhat misleading as suggested by \cref{fig:phi4-comparison-LFRG-MC-2d}. 
There, our results for the magnetization and susceptibility are shown as a function of~$m^2$. 
From this figure, we deduce that the results obtained with the two methods do not agree in the vicinity of the phase transition. 

In \cref{fig:phi4-comparison-LFRG-MC-2d-Ns-scan-critical-regime}, we show the magnetization and susceptibility as a function of the external field~$c$ again, but now for~$m^2=m^2_{\mathrm{peak}} \approx -1.295$ (vertical dashed line in \cref{fig:phi4-comparison-LFRG-MC-2d}).
At this value of~$m^2$, we have the greatest deviation of the lattice \gls{FRG} and lattice \gls{MC} results in \cref{fig:phi4-comparison-LFRG-MC-2d}, in accordance with our \gls{FRG}-intrinsic analysis of the predictive power of \gls{LPA} in \cref{sec:assessing-lpa}. 
We observe in \cref{fig:phi4-comparison-LFRG-MC-2d-Ns-scan-critical-regime} that the lattice \gls{FRG} and lattice \gls{MC} results for~$m^2= m^2_{\mathrm{peak}}$ deviate from each other over a wide range of external field values, down to smaller and smaller values of~$c$ as~$N_{\sigma}$ increases. 
In particular, we find that the deviations already appear on comparatively small lattices. 
For sufficiently large values of~$c$, the results from the two methods are in good agreement. 
However, this is not surprising: fluctuation effects and momentum dependences in correlation functions are suppressed in this regime since the mass of the scalar field increases with~$c$. 

Recall that the magnetization as a function of the external field is directly related to the field derivative of the effective potential as a function of the field, see \cref{subsubsec:effpotcomp}. 
Thus, the deviations of the lattice \gls{FRG} results for the magnetization at small~$c$ from the lattice \gls{MC} results would translate into corresponding deviations in the predictions for the effective potential near its minimum and at small field values.

We emphasize that the deviations in the results from the two methods are (strongly) suppressed (far) away from the critical region in two spacetime dimensions, see, e.g., \cref{fig:phi4-comparison-LFRG-MC-2d}. 
In any case, the deviations are generally much smaller in three spacetime dimensions, even near the phase transition, see, e.g., \cref{fig:phi4-comparison-LFRG-MC-3d}.

\vfill

\bibliography{bib}
\end{document}